\documentclass[11pt]{article}
\pdfoutput=1
\usepackage[centertags]{amsmath}
\usepackage[square, comma, sort&compress,numbers]{natbib}
\usepackage{array,multirow}
\numberwithin{equation}{section}
\usepackage{amssymb}
\usepackage{graphicx}
\usepackage{color}

\usepackage{epsfig}
\usepackage{bbold}

\usepackage{wrapfig}
\usepackage{float}
\usepackage{soul}

\usepackage{tikz,pgf}
\usetikzlibrary{shapes}
\usetikzlibrary{calc}
\usetikzlibrary{decorations.pathmorphing}
\usetikzlibrary{decorations.pathreplacing,shapes.misc}
\usetikzlibrary{positioning}
\usetikzlibrary{arrows}
\usetikzlibrary{decorations.markings}
\usetikzlibrary{shadings}

\usetikzlibrary{intersections}

\def\be{\begin{equation}}
\def\ee{\end{equation}}
\def\ba{\begin{array}}
\def\ea{\end{array}}

\def\dps{\displaystyle}
\newcommand{\half}{\frac{1}{2}}
\def\tr{{\rm Tr}}

\def\1{\tilde{1}}
\def\2{\tilde{2}}
\def\3{\tilde{3}}

\newdimen\tableauside\tableauside=1.0ex
\newdimen\tableaurule\tableaurule=0.4pt
\newdimen\tableaustep
\def\phantomhrule#1{\hbox{\vbox to0pt{\hrule height\tableaurule
width#1\vss}}}
\def\phantomvrule#1{\vbox{\hbox to0pt{\vrule width\tableaurule
height#1\hss}}}
\def\sqr{\vbox{%
  \phantomhrule\tableaustep

\hbox{\phantomvrule\tableaustep\kern\tableaustep\phantomvrule\tableaustep}%
  \hbox{\vbox{\phantomhrule\tableauside}\kern-\tableaurule}}}
\def\squares#1{\hbox{\count0=#1\noindent\loop\sqr
  \advance\count0 by-1 \ifnum\count0>0\repeat}}
\def\tableau#1{\vcenter{\offinterlineskip
  \tableaustep=\tableauside\advance\tableaustep by-\tableaurule
  \kern\normallineskip\hbox
    {\kern\normallineskip\vbox
      {\gettableau#1 0 }%
     \kern\normallineskip\kern\tableaurule}%
  \kern\normallineskip\kern\tableaurule}}
\def\gettableau#1 {\ifnum#1=0\let\next=\null\else
  \squares{#1}\let\next=\gettableau\fi\next}

\newcommand{\bref}[1]{\textbf{\ref{#1}}}

\def\cA{\mathcal{A}}
\def\cB{\mathcal{B}}

\def\cF{\mathcal{F}}

\def\cO{\mathcal{O}}

\def\cV{\mathcal{V}}

\numberwithin{equation}{section} \makeatletter
\@addtoreset{equation}{section}

\def\be{\begin{equation}}
\def\ee{\end{equation}}
\def\ba{\begin{array}}
\def\ea{\end{array}}

\def\dps{\displaystyle}

\def\ba{\begin{array}}
\def\ea{\end{array}}

\def\dps{\displaystyle}

\def\cl{c\to\infty}

\usepackage{jheppub}

\makeatletter
\def\@fpheader{\vspace{-.1cm}}
\makeatother

\title{Holographic duals of large-$c$  torus conformal  blocks}

\author[a,b]{Konstantin\ Alkalaev,}

\author[a,c]{Vladimir\ Belavin}

\affiliation[a]{I.E. Tamm Department of Theoretical Physics, \\P.N. Lebedev Physical
Institute,\\ Leninsky ave. 53, 119991 Moscow, Russia}
\affiliation[b]{Department of General and Applied Physics, \\
Moscow Institute of Physics and Technology, \\
Institutskiy per. 7, Dolgoprudnyi, \\141700 Moscow region, Russia}
\affiliation[c]{Department of Quantum Physics, \\
Institute for Information Transmission Problems, \\
Bolshoy Karetny per. 19, 127994 Moscow, Russia}
\emailAdd{alkalaev@lpi.ru}
\emailAdd{belavin@lpi.ru}

\abstract{ We study CFT$_2$ conformal blocks on a torus and their holographic realization.  
The  classical conformal blocks arising in the regime where conformal dimensions grow linearly with the large central charge are shown to be holographically dual to the geodesic networks stretched in the thermal AdS bulk space. We discuss the $n$-point conformal blocks and their duals,  the 2-point case is elaborated in full detail.  We develop various techniques to calculate both quantum and classical conformal block functions. In particular, we show that exponentiated global torus blocks reproduce classical torus blocks in the specific perturbative regimes of the conformal parameter space.  }

\keywords{CFT, Virasoro algebra, torus, AdS/CFT duality}
\preprint{FIAN-TD-2017-20}
\arxivnumber{}

\begin{document}

\maketitle
\flushbottom

\section{Introduction}  

The study of the AdS$_3$/CFT$_2$ correspondence in the large central charge approximation has lead to the simple formula relating the classical conformal block $f_{class}$  and the length  of  geodesic network $L_{dual}$  in the dual spacetime 
\be
\label{blf}
f_{class}(\epsilon, \tilde \epsilon |z) \cong L_{dual}(\epsilon, \tilde \epsilon|w)\;,
\ee
where coordinates $z = z(w)$ are given by the conformal map from the AdS boundary to the plane CFT, $\epsilon, \tilde\epsilon$ are conformal dimensions, the equality symbol $\cong$ means that both sides are equal modulo logarithmic terms defined by the conformal map \cite{Hartman:2013mia,Fitzpatrick:2014vua,Asplund:2014coa,Caputa:2014eta,deBoer:2014sna,Hijano:2015rla,Fitzpatrick:2015zha,Alkalaev:2015wia} (for the further development see \cite{Hijano:2015qja,Alkalaev:2015lca,Beccaria:2015shq,Fitzpatrick:2015dlt,Banerjee:2016qca,Chen:2016dfb,Alkalaev:2016ptm,Kraus:2016nwo,Hulik:2016ifr,Fitzpatrick:2016mtp,Kraus:2017ezw,Belavin:2017atm}).  The large-$c$ conformal blocks can be defined by  a number of heavy or light operators with dimensions growing linearly with $c$. The corresponding bulk geometry can be either conical singularity/BTZ \cite{Fitzpatrick:2014vua} or thermal AdS \cite{Alkalaev:2016ptm} in the case of the boundary spherical or toric CFT, respectively.

In this paper we continue to study large-$c$ CFT$_2$ on a torus from the holographic perspective. Basically, we consider the $2$-point torus blocks  generalizing to the $n$-point case wherever possible. We  calculate the corresponding classical blocks within various approximations and show that they are holographically realized as geodesic networks on the thermal AdS in keeping with the block/length correspondence formula \eqref{blf}.

The outline of the paper is as follows. In Section \bref{sec:n-point} we first  shortly discuss  $n$-point correlation functions on a torus and then focus on  the $n=2$ case.  In order to calculate the block functions we develop two methods, (i)  a straightforward evaluation of the constituent matrix elements, (ii)   a combinatorial (AGT) representation, see Appendices \bref{app:matr}, \bref{app:t}, and \bref{sec:AGT}, respectively.\footnote{These results are complementary to the recursive representations of 2-point torus blocks  \cite{Cho:2017oxl}.} 

In Section \bref{sec:class}--\bref{sec:glob} we discuss various limiting  torus blocks paying particular attention to the so called classical and global blocks.\footnote{Global blocks and their holographic duals were recently discussed in  \cite{Kraus:2017ezw}.} In Section \bref{sec:exp} we discuss the {\it classical global} torus blocks that arise as exponentiation of the global  blocks in the regime of large conformal dimensions. We also show that classical global blocks describe the linearized part of the standard classical blocks within particular perturbation theory when some conformal dimensions are larger than the others. This yields the method to calculate approximate classical blocks that bypass the  full quantum Virasoro block analysis.   

The holographic interpretation of  classical torus blocks is discussed in Section  \bref{sec:dual}. Here we propose the general scheme and formulate the system of differential and algebraic equations that describes the dual network. However, similar to the $n$-point sphere case, exact solutions to the equation system  are not known yet. Instead, in Section \bref{sec:sat} we propose a particular perturbation theory and find approximate solutions in the $2$-point case.   

We close with a brief conclusion in Section \bref{sec:discussion}. All technicalities and subsidiary discussions  are collected in Appendices \bref{app:matr} -- \bref{sec:first}.

\section{Torus  conformal blocks}
\label{sec:n-point}

An $n$-point torus correlation function of arbitrary primary operators $\phi_i(z_i, \bar z_i)$  with conformal dimensions $(\Delta_i, \bar \Delta_i)$  is given by 
\be
\label{2pt}
\langle \phi_1(z_1, \bar z_1) \cdots \phi_n(z_n, \bar z_n) \rangle = (q\bar q)^{-\frac{c}{24}}\,\tr\left(q^{L_0}\bar q^{\bar L_0} \phi_1(z_1, \bar z_1) \cdots \phi_n(z_n, \bar z_n)  \right)\;,
\ee
where $q = e^{2\pi i \tau_{cft}}$,  $\tau_{cft}\in \mathbb{C}$ is the torus modular parameter, $L_0$ and $\bar L_0$ are the Virasoro generators, $c$ is the central charge. Since a torus can be  realized as a cylinder with the edges rotated and glued together, the torus correlation functions can be defined using the plane CFT notation (we always assume that \eqref{2pt} is supplemented by  the conformal map from the plane to the cylinder) \cite{Cardy:1986ie}. Assuming that the space of states is generated by primary operators with dimensions denoted as $(\tilde \Delta_1, \bar{\tilde\Delta}_1)$ and evaluating the trace we find  that the torus correlation functions is given by a power series  in the modular parameter,    
$$
\langle \phi_1(z_1, \bar z_1) \cdots \phi_n(z_n, \bar z_n) \rangle = (q\bar q)^{-\frac{c}{24}}\sum_{(\tilde \Delta_1, \bar{\tilde \Delta}_1)}\;\sum_{m=0}^\infty q^{\tilde \Delta_1+m} \bar q^{\bar{\tilde\Delta}_1+m}\times 
$$
\be
\label{npt_dec}
\times \sum_{m = |M|=|N|}\sum_{m = |\bar M|=|\bar N|}  B^{M|N} B^{\bar M| \bar N} \;
F(\Delta_i,\bar \Delta_i, \tilde \Delta_1, \bar{\tilde \Delta}_1, M,N,  \bar M, \bar N|z_i, \bar z_i)\;,
\ee
where the matrix element of $n$ primary operators  
\be
\label{Fmatr}
F(\Delta_i,\bar \Delta_i, \tilde \Delta_1, \bar{\tilde \Delta}_1, M,N,  \bar M, \bar N|z, \bar z)
 = \langle  \tilde \Delta_1, \bar{\tilde \Delta}_1, M, \bar M  |\phi_1(z_1, \bar z_1) \cdots \phi_n(z_n, \bar z_n)|N, \bar N, \tilde \Delta_1, \bar{\tilde \Delta}_1\rangle\;,
\ee
is given  in the standard basis   
$
|M, \bar M, \tilde\Delta_1, \bar{\tilde \Delta}_1 \rangle = \bar L_{-n_1}^{j_1} \cdots  \bar L_{-n_l}^{j_l}L_{-m_1}^{i_1} \cdots  L_{-m_k}^{i_k} |\tilde\Delta_1, \bar{\tilde \Delta}_1\rangle \;.
$ 
Here, descendant vectors in the Verma module are generated from the primary state $|\tilde\Delta_1, \bar{\tilde \Delta}_1\rangle$, indices  $M, \bar M$ label basis monomials, $|M| = i_1m_1+ \ldots +i_k m_k$ and $|\bar M| = j_1 n_1+ \ldots +j_l n_l$. The matrix $B^{M|N}$ is the inverse of the Gram matrix.

In what follows we focus on the $n=2$ point case and expand  the matrix elements \eqref{Fmatr} into conformal blocks. Different exchanged channels can be obtained by plugging  resolutions of identity  into the matrix element \eqref{Fmatr} and/or using the OPE.  Note that conformal dimensions $(\tilde\Delta_1, \bar{\tilde \Delta}_1)$ associated to the $q$-expansion \eqref{npt_dec} partially describe possible exchanged channels and, in fact, define an (un)closed loop part of the corresponding diagrams.

\begin{figure}[H]
\centering

\begin{tikzpicture}[line width=1pt,scale=0.60]

%%%%%%%%%%%%% the first  graph %%%%%%%%%%%%%%

\draw (-10,-1) -- (-8,-1);
\draw (-10,-1) -- (-11,-2);
\draw (-10,-1) -- (-11,0);
\draw (-8,-1) -- (-7,-2);
\draw (-8,-1) -- (-7,0);

\draw[blue, dashed] (-11,0) .. controls (-13,3) and (-5,3) .. (-7,0);

%%%%%%%%%%%%% the second graph %%%%%%%%%%%%%%

\draw (0,0) -- (2,0);
\draw (0,0) -- (-1,-1);
\draw (0,0) -- (-1,1);
\draw (2,0) -- (3,-1);
\draw (2,0) -- (3,1);

\draw[blue, dashed] (3,1) .. controls (6,3) and (6,-3) .. (3,-1);

\end{tikzpicture}

\caption{ Two-point conformal blocks in the $s$-channel (right) and $t$-channel (left). The $s$-channel loop consists of two segments with different conformal dimensions.} 
\label{24ch}
\end{figure}
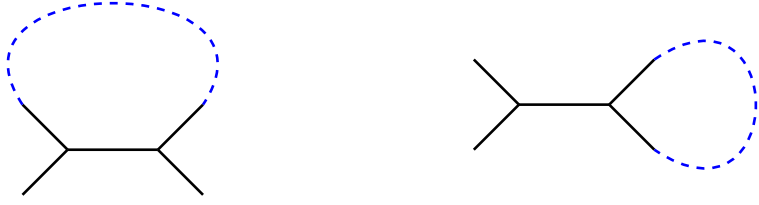

In the 2-point case, the  torus correlation functions can be expanded in two channels that we call $s$-channel and $t$-channel, see Fig. \bref{24ch}. They directly follow from \eqref{2pt}: the trace over the space of states can be understood as a sum of 4-point correlation functions on a sphere over two outermost descendant operators in points $\infty$ (left) and $0$ (right). Then, using the OPE and recalling that on a sphere there are three different 4-point exchanged channels we can see that identifying  two external legs we get just two topologically non-equivalent configurations, Fig. \bref{24ch}.

\vspace{-5mm}

\paragraph{$s$-channel.} Two-point torus conformal block in the $s$-channel (left diagram on Fig. \bref{24ch}) is defined by  inserting the resolution of identity between two primary operators in the matrix element \eqref{Fmatr},
\be
\mathbb{1} = \sum_{\Delta, \bar \Delta}\;\,\sum_{m=0}^\infty\; \sum_{|S|=|T|=m}\sum_{|\bar S| = |\bar T|=m} |S,\bar S,\Delta, \bar \Delta\rangle\, B^{S|T}B^{\bar S|\bar T}\,\langle \Delta,\bar \Delta, T , \bar T|\;,
\ee 
where the inverse Gram matrix $B^{S|T}$ enters the formula because the standard basis is non-diagonal. Then  \eqref{Fmatr} splits into products of two 3-point functions of primary operator with  two descendant operators. Recalling that  $\langle \tilde \Delta_m|\phi_k(z_k)|\tilde\Delta_l\rangle =  C_{\tilde\Delta_m\Delta_k\tilde\Delta_k}z_k^{\tilde\Delta_m-\Delta_k-\tilde\Delta_l}$, where $C_{\tilde\Delta_m\Delta_k\tilde\Delta_k}$ are the structure constants,  
we decompose the 2-point correlation function as follows
\be
\ba{l}
\dps
\langle \phi_1(z_1, \bar z_1)  \phi_2(z_2, \bar z_2) \rangle = \sum_{(\tilde \Delta_1, \bar{\tilde \Delta}_1)}\sum_{(\tilde \Delta_2, \bar{\tilde \Delta}_2)}C_{\tilde\Delta_1\Delta_1\tilde\Delta_2}C_{\tilde\Delta_2\Delta_2\tilde\Delta_1}\times
\\
\\
\dps
\hspace{0mm}\times z_1^{\tilde\Delta_1-\Delta_1-\tilde\Delta_2}\bar z_1^{\bar{\tilde\Delta}_1-\bar\Delta_1-\bar{\tilde\Delta}_2}z_2^{\tilde\Delta_2-\Delta_2-\tilde\Delta_1}\bar z_2^{\bar{\tilde\Delta}_2-\bar \Delta_2-\bar{\tilde\Delta}_1}\left[\cV_c^{\Delta_{1,2},\tilde\Delta_{1,2}}(q, z_{1,2})\cV_c^{\bar\Delta_{1,2}, \bar{\tilde \Delta}_{1,2}}(\bar q, \bar z_{1,2})\right]\;,
\ea
\ee
where the $s$-channel (holomorphic) conformal block is \footnote{Differently parameterized Virasoro torus blocks were considered in the $2$-point case in Ref. \cite{Brustein:1988vb}, in the $n$-point case in Ref.  \cite{Cho:2017oxl}, and for minimal models in Ref. \cite{Jayaraman:1988ex}. For a related discussion of  $n$-point global blocks see \cite{Kraus:2017ezw}.}

\be
\label{necklace}
\ba{l}
\dps
\cV_c^{\Delta_{1,2},\tilde\Delta_{1,2}}(q, z_{1,2})=
\\
\\
\dps
\hspace{8mm}= q^{c/24 - \tilde\Delta_1}\sum_{n,m=0}^\infty q^{n}\sum_{\substack{|M|=|N|=n\\|S|=|T|=m}} B^{M|N}_1\;
\frac{\langle \tilde \Delta_1, M |\phi_1(z_1)|S,\tilde\Delta_2\rangle}{\langle \tilde \Delta_1|\phi_1(z_1)|\tilde\Delta_2\rangle}\; B^{S|T}_2\;
\frac{\langle \tilde \Delta_2, T |\phi_2(z_2)|N,\tilde\Delta_1\rangle}{
\langle \tilde \Delta_2 |\phi_2(z_2)|\tilde\Delta_1\rangle}\;,

\ea
\ee
where $\tilde \Delta_{1,2}$ are exchanged conformal dimensions. Given that  all inner products in \eqref{necklace} have been explicitly calculated we arrive at the double power series in the modular parameter $q$ and the ratio $x = z_2/z_1$ with the expansion coefficients being rational functions of  conformal parameters,
\be
\label{sblock}
\cV_c^{\Delta_{1,2},\tilde\Delta_{1,2}}(q, x) = q^{c/24 - \tilde\Delta_1}\sum_{n\in \mathbb{N}_0}\; \sum_{m\in \mathbb{Z}} \; \cV^{\Delta_{1,2},\tilde\Delta_{1,2}}_{c\, (n,m)}\, q^{n}\, x^m\;.
\ee 
Using explicit formulas from Appendix \bref{app:matr} we write down the block function in the lowest orders,\footnote{In principle, the matrix approach can be used to find coefficients in any order. Instead, in Appendix \bref{sec:AGT} we develop the {\it combinatorial representation} of the $n$-point conformal torus blocks in the $s$-channel. Using Mathematica we apply this representation to compute the block coefficients up to high enough order in $q$ and $x$. 
}
\be
\label{schan}
\ba{l}
\dps
\cV_c^{\Delta_{1,2},\tilde\Delta_{1,2}}(q, x) =
\\
\\
\dps
= \Big[1+ \frac{(\Delta_1 +\tilde\Delta_2 -\tilde \Delta_1)(\Delta_2 +\tilde\Delta_2 -\tilde \Delta_1)}{2\tilde\Delta_2}x+ \cO(x^2)\Big]
 + q \Big[\frac{(\tilde \Delta_1 +\Delta_1 -\tilde \Delta_2)(\tilde \Delta_1 +\Delta_2 -\tilde \Delta_2)}{2\tilde\Delta_1} x^{-1}
\\
\\
\dps
+ \frac{\left((\tilde\Delta_2+\Delta_1-\tilde\Delta_1)(\tilde\Delta_1+\Delta_1-\tilde\Delta_2-1)+2\tilde\Delta_2\right)\left((\tilde\Delta_1+\Delta_2-\tilde\Delta_2)(\tilde\Delta_2+\Delta_2-\tilde\Delta_1-1)+2\tilde\Delta_1\right)}{4\tilde\Delta_1\tilde\Delta_2} \\
\\
\\
+\cO(x^1)\Big] +\cO(x^m,q^k)\;.
\\
\ea
\ee
Setting $\Delta_2 = 0$, $\Delta_1 \equiv \Delta$ and  equating $\tilde\Delta_1 = \tilde\Delta_2 \equiv \tilde \Delta$ we  reproduce the 1-point torus block \cite{Hadasz:2009db} with external dimension $\Delta$ and exchanged  dimension $\tilde \Delta$. From the form of coefficients in \eqref{schan} it follows that {\it vacuum} ($\tilde \Delta_1 =0$ or $\tilde\Delta_2 = 0$) blocks in this channel are absent.  

\vspace{-1mm}

\paragraph{$t$-channel.} Alternatively, we can use the OPE of two primary operators in the matrix element \eqref{Fmatr}. In this case \eqref{Fmatr} reduces to summing over  3-point functions of three descendant operators with the OPE coefficients. Namely, we fuse two primaries    
\be
\label{OPE}
\phi_1(z_1)\phi_2(z_2) = \sum_{\tilde\Delta_2} C_{_{\Delta_1\Delta_2\tilde\Delta_2}} (z_1-z_2)^{\tilde\Delta_2 - \Delta_1-\Delta_2} \psi_{\tilde\Delta_2}(z_1,z_2)\;,
\ee 
where the resulting  operator  is given by 
\be
\label{OPEpsi}
\psi_{\tilde\Delta_2}(z_1,z_2) = \phi_{\tilde\Delta_2}(z_2) + \beta_1 (z_1-z_2) L_{-1} \phi_{\tilde\Delta_2}(z_2)+ ...\;, 
\qquad 
\beta_1 = \frac{\tilde\Delta_2+\Delta_1 - \Delta_2}{2\tilde\Delta_2}\;.
\ee
Plugging the OPE \eqref{OPE} into the the correlation function \eqref{npt_dec} we find
\be
\ba{l}
\dps
\langle \phi_1(z_1, \bar z_1)  \phi_2(z_2,\bar z_2) \rangle = (q\bar q)^{-\frac{c}{24}} \sum_{(\tilde\Delta_1, \bar{\tilde\Delta}_1)}\sum_{(\tilde\Delta_2, \bar{\tilde\Delta}_2)} C_{_{\Delta_1\Delta_2\tilde\Delta_2}}C_{_{\bar{\Delta}_1\bar{\Delta}_2\bar{\tilde\Delta}_2}} \times
\ea
\ee
$$
\dps
\times (z_1-z_2)^{\tilde\Delta_2 - \Delta_1-\Delta_2} \;(\bar z_1-\bar z_2)^{\bar{\tilde\Delta}_2 - \bar\Delta_1-\bar\Delta_2}\;z_2^{-\tilde\Delta_2} \bar z_2^{-\bar{\tilde\Delta}_2}\; \left[\cV_c^{\Delta_{1,2},\tilde\Delta_{1,2}}(q, z_{1,2})\cV_c^{\bar\Delta_{1,2}, \bar{\tilde \Delta}_{1,2}}(\bar q, \bar z_{1,2})\right]\;,
$$
where the $t$-channel (holomorphic) conformal block is
\be
\label{stickman}
\ba{l}
\dps
\cV_c^{\Delta_{1,2},\tilde\Delta_{1,2}}(q, z_{1,2}) = 
 \sum_{n=0}^\infty q^{n}\sum_{|M|=|N|=n} B^{M|N}_1 
\frac{\langle \tilde \Delta_1, M |\psi_{\tilde\Delta_2}(z_1,z_2)|N,\tilde\Delta_1\rangle}{\langle \tilde \Delta_1|\phi_{\tilde\Delta_2}(z_2)|\tilde\Delta_1\rangle}\;.

\ea
\ee
where $\tilde \Delta_{1,2}$ are exchanged conformal dimensions. Given that 3-point functions of three descendants in \eqref{stickman} have been explicitly calculated we arrive at the double power series  in the modular parameter $q$ and the ratio $w = (z_1-z_2)/z_2$ with the expansion coefficients being rational functions of  conformal parameters,
\be
\cV_c^{\Delta_{1,2},\tilde\Delta_{1,2}}(q, w) = q^{c/24 - \tilde\Delta_1}\sum_{n\in \mathbb{N}_0}\; \sum_{m\in \mathbb{N}_0} \;   \cV_{c\,(n,m)}^{\Delta_{1,2}, \tilde \Delta_{1,2}}q^{n}\, w^m\;.
\ee 
Note that the $t$-channel block  is the power series in $w$-variable, while the $s$-channel block is the Laurent series in $x$-variable, cf. \eqref{sblock}. Using explicit formulas from Appendix \bref{app:t} we write down the block function in the lowest orders,
\be
\label{tchan}
\ba{l}
\dps
\cV_c^{\Delta_{1,2},\tilde\Delta_{1,2}}(q, w) =\Big[1-\frac{\tilde\Delta_2+\Delta_1 - \Delta_2}{2}w+ \cO(w^2)\Big]+ q \Big[\frac{2\tilde\Delta_1 +\tilde\Delta_2(\tilde\Delta_2-1)}{2\tilde\Delta_1} \\
\\
\dps
\hspace{45mm}-\frac{(2 \tilde\Delta_1+(\tilde\Delta_2-1) \tilde\Delta_2) (\Delta_1+\tilde\Delta_2-\Delta_2)}{4 \tilde\Delta_1}w+\cO(w^2)\Big] \;.
\ea
\ee 

The 1-point block with dimensions $\Delta, \tilde\Delta$ is reproduced by setting $\Delta_1=0$, $\Delta_2 = \tilde\Delta_2 \equiv \Delta$, and $\tilde \Delta_1 \equiv \tilde \Delta$. 
Let us note that there exist {\it vacuum} $t$-channel conformal blocks which arise when the second exchanged operator (an intermediate straight line on the right diagram on Fig. \bref{24ch}) is the unity operator. Thus, setting $\tilde\Delta_2 = 0$ supplemented by the fusion condition $\Delta_1 = \Delta_2\equiv\Delta$ we find the vacuum block function  which depends on the external dimension $\Delta$, and the loop exchanged dimension $\tilde \Delta$.

\section{Classical two-point torus  blocks }
\label{sec:class}

The parameter space of the conformal block  functions  $\cV^{\Delta, \tilde\Delta}_c(q|z)$ includes external and intermediate  dimensions,  modular parameters, and the central charge. In this section we shall discuss {\it semiclassical} blocks that correspond to different asymptotics in the parameter space when the conformal dimensions scale  differently with the central charge. Near the point $c = \infty$ we distinguish between heavy and light dimensions depending on how a given dimension scales with the central charge: $\Delta_{light} \approx \epsilon$ or  $\Delta_{heavy} \approx c \epsilon $, where $\epsilon$ are classical dimensions.  Assuming that a conformal block depends on a number of light and heavy operators and expanding around $c = \infty$ we find that 
\be
\label{laurent}
\cV^{\Delta, \tilde\Delta}_{c}(q,z)\, \cong\,   \sum_{n\in \mathbb{N}} \frac{v_n^{\epsilon, \tilde \epsilon}(q,z)}{c^n}\;,
\ee
where the expansion coefficients are power series in the modular parameter $q$ with coefficients being rational functions of classical dimensions $\epsilon$, $\tilde\epsilon$. 

The principal part of \eqref{laurent} vanishes in the large-$c$ limit, while the form  of the regular part defines a particular semiclassical block. It is known that there are different types of semiclassical blocks, including the light blocks, various heavy-light blocks, and the classical block (see e.g. discussion in \cite{Fitzpatrick:2015zha,Alkalaev:2015fbw,Alkalaev:2016fok,Cho:2017oxl}). \footnote{Generally speaking, global torus blocks cannot be obtained in this way, see Section \bref{sec:glob}.        
} Using the expansion \eqref{laurent} we see that e.g. the light block is given by  $v_0^{\epsilon, \tilde \epsilon}(q,z)$, while other coefficients vanish, $v_n^{\epsilon, \tilde \epsilon}(q,z) =0$, $n>0$. The classical block has non-vanishing coefficients $v_n^{\epsilon, \tilde \epsilon}(q,z)\neq 0$, $\forall n \in \mathbb{N}_0$. In particular, the regular part of the classical block is claimed to be an exponential functional  linear in $c$.

%
%Note that  the global blocks on any Riemann  surface except for a sphere  cannot be obtained as the large $c$ limit of the full Virasoro block with light operators \cite{Alkalaev:2016fok,Cho:2017oxl}: the limiting function here  is the light block. In the sphere case, the global and light blocks coincide. We shall discuss global torus block in Section \bref{sec:glob}.      

In what follows we focus on the classical torus block which therefore is given by 
\be
\label{ccb}
\cV^{\Delta_{1,2}, \tilde \Delta_{1,2}}_{c}(q, z_{1,2} ) \, \cong\,  \exp\big[\,\frac{c}{6}\,f^{\epsilon_{1,2}, \tilde \epsilon_{1,2}}(q, z_{1,2})\big]\qquad  \text{as}\qquad  \cl\;,
\ee
where function $f^{\epsilon_{1,2}, \tilde \epsilon_{1,2}}$ is the corresponding classical conformal block conveniently parameterized by the classical conformal dimensions 
\be
\label{deltas}
\epsilon_i = \frac{\Delta_i}{k}\,  \;,\quad\;\; \tilde\epsilon_i = \frac{\tilde \Delta_i}{k} \;,\qquad\text{where}\qquad k=\frac{c}{6}\;.
\ee

\paragraph{s.} Using \eqref{schan} we find the $s$-channel block function    
\be
\label{class_block_1}
\ba{l}
\dps
f^{\epsilon_{1,2}, \tilde \epsilon_{1,2}}(q, x ) = (\tilde\epsilon_1 - 1/4) \log q + \sum_{n=0}^\infty q^n \text{f}^{\,(1)}_n (\epsilon, \tilde \epsilon|x)\;,
\ea
\ee
where a few lowest level coefficients are given by 
\be
\label{fir}
\ba{l}
\dps
\text{f}^{\,(1)}_0 (\epsilon, \tilde \epsilon|x) = \frac{(\epsilon_1 +\tilde\epsilon_2 -\tilde \epsilon_1)(\epsilon_2 +\tilde\epsilon_2 -\tilde \epsilon_1)}{2\tilde\epsilon_2} \,x + \cO(x^2)\;,
\\
\\
\dps
\text{f}^{\,(1)}_1 (\epsilon, \tilde \epsilon|x) = \frac{(\epsilon_1+\tilde \epsilon_1 -\tilde \epsilon_2)(\epsilon_2+\tilde \epsilon_1 -\tilde \epsilon_2)}{2\tilde\epsilon_1}\,x^{-1}+ \cO(x^0)\;.
\ea
\ee

\paragraph{t.} Using \eqref{tchan} we find the $t$-channel block function 
\be
\label{class_block_2}
\ba{l}
\dps
f^{\epsilon_{1,2}, \tilde \epsilon_{1,2}}(q, w )  = (\tilde\epsilon_1 - 1/4) \log q + \sum_{n=0}^\infty q^n \text{f}^{\,(2)}_n (\epsilon, \tilde \epsilon|w)\;,\ea
\ee
where a few lowest level coefficients are given by 
\be
\ba{l}
\dps
\text{f}^{\,(2)}_0 (\epsilon, \tilde \epsilon|w) = -\frac{\left(\tilde{\epsilon }_2+\epsilon _1-\epsilon _2\right)}{2}\,w+\cO(w^2)\;,
\qquad\quad 
\text{f}^{\,(2)}_1 (\epsilon, \tilde \epsilon|w) = \frac{\tilde{\epsilon }_2^2}{2 \tilde{\epsilon }_1} +\cO(w)\;.
\ea
\ee

\section{Perturbative  classical  $s$-channel  torus blocks}
\label{sec:heavy-light}

The analysis of  classical blocks simplifies within various approximations where some of dimensions are much larger than the others, see, e.g. \cite{Fitzpatrick:2014vua,Hijano:2015rla,Alkalaev:2015wia,Banerjee:2016qca,Alkalaev:2016ptm}. In the torus case, already from the first coefficients \eqref{fir} (for simplicity, we choose pairwise equal dimensions) $\text{f}_0 = \frac{\epsilon_1^2}{2\tilde\epsilon_1}$  we immediately conclude  that  it blows up when  $\tilde\epsilon_1 \ll \epsilon_1$ and is smooth in the opposite regime $\epsilon_1 \ll \tilde\epsilon_1$. In what follows we study the torus $s$-channel blocks assuming that the exchanged channels with equal dimensions $\tilde \epsilon \equiv \tilde \epsilon_1 = \tilde \epsilon_2$ are much heavier than the external operators $\epsilon_1$ and $\epsilon_2$, i.e. $\epsilon_i/\tilde\epsilon \ll 1$. This step provides the first level perturbation expansion because the external dimensions remain unrelated to each other. On the second level, we find two possible perturbation expansions. In the first version called the superlight expansion \cite{Alkalaev:2015wia,Alkalaev:2015lca,Alkalaev:2015fbw} we assume that the second external dimension is much less than the first one, $\epsilon_1/\epsilon_2\ll1$. In the second version that we call the double leg expansion the external dimensions are equated.

\paragraph{Superlight expansion.} In this case the dimensions are restricted as
\be
\label{constraints1}
\tilde\epsilon \equiv \tilde\epsilon_1 = \tilde\epsilon_2\;,
\qquad
\delta  \equiv \frac{\epsilon_1}{\tilde\epsilon} \ll 1\;,
\qquad
\nu  \equiv \frac{\epsilon_2}{\epsilon_1} \ll 1\;.
\ee
The first constraint is the fusion rule guaranteeing that   $\epsilon_1=0$ and/or $\epsilon_2=0$ are consistent.  The corresponding conformal block is parameterized by three parameters  $\tilde \epsilon$, $\delta$, $\nu$ so that we arrive at the triple deformation theory:  we expand around $\tilde\epsilon = \infty$ and then around $\delta=0$ and $\nu = 0$. Keeping terms linear in $\tilde\epsilon$ we arrive at the quartic series expansion  
\be
\label{linblock}
f_{\rm{lin}}^{\tilde\epsilon,\delta,\nu}(q, x)\equiv (\tilde\epsilon - 1/4) \log q + \tilde \epsilon \sum_{n=1}^\infty f^{(1)}_{2n} \delta^{2n} +  \tilde\epsilon \nu \sum_{n=2}^\infty f^{(2)}_n\delta^n + \cO(\tilde\epsilon^2, \nu^2)\;,
\ee
where the second term is the perturbative 1-point block with coefficients \cite{Alkalaev:2016ptm,Alkalaev:2016fok} 
\be
\label{f2f4}
f^{(1)}_{2} = \frac{q}{2}\frac{1}{1-q}\;,
\qquad
f^{(1)}_{4} =-\frac{q^2}{48}\frac{q-3}{(1-q)^3}\;,\qquad ...\;\;\;, 
\ee
because setting $\nu=0$ we automatically reproduce the 1-point case, and the third term gives the $\nu$-correction, where a few first coefficients read
\be
f^{(2)}_2 = \frac{x}{2}+\left(\frac{x}{2}+\frac{1}{2x}\right) q+... \;,
\qquad
f^{(2)}_3 = \frac{x^2}{8}+ \left(\frac{x^2}{4}-\frac{1}{4}\right)q +...\;.
\ee
More higher order terms can be found in \eqref{bloook}.

\paragraph{Double leg expansion.} The conformal dimensions are pairwise equal and satisfy the constraints
\be
\label{constraints2}
\tilde\epsilon_1 = \tilde\epsilon_2\equiv \tilde \epsilon\;,
\qquad
\epsilon_1 = \epsilon_2\equiv \epsilon\;,
\qquad
\delta  \equiv \frac{\epsilon}{\tilde\epsilon} \ll 1\;.
\ee
The corresponding block depends on two parameters $\tilde\epsilon$ and $\delta$ and, therefore, we consider the double perturbation theory: we expand around $\tilde\epsilon = \infty$ and then around $\delta=0$. Keeping terms linear in $\tilde\epsilon$ we arrive at the triple series expansion  
\be
\label{linblock}
f_{\rm{lin}}^{\tilde\epsilon,\delta}(q, x)\equiv (\tilde\epsilon - 1/4) \log q + \tilde \epsilon \sum_{n=2}^{\infty} g_{n}(q,x) \delta^{n}+\cO(\tilde\epsilon^2, \nu^2)\;,
\ee
where a few first coefficients read 
\be
\label{firstlincoef}
g_{2}(x,q) = \frac{x}{2}+q \left(\frac{x}{2}+\frac{1}{2 x}+1\right)+...\;,
\qquad
g_{3}(x,q) =\frac{x^2}{4}+q \left(\frac{x^2}{2}-\frac{1}{2}\right)+...\;.
\ee
The higher order expression can be found in \eqref{classblockhigh}.

\section{Global torus blocks}
\label{sec:glob}

Global blocks of CFT$_2$ are associated to $sl(2, \mathbb{R}) \subset Vir$. Equivalently, the global blocks can be obtained by considering particularly contracted Virasoro algebra at $c\to\infty$ while keeping both external and exchanged conformal dimensions independent of the central charge, $\Delta_i=\cO(c^0)$ and $\tilde \Delta_j=\cO(c^0)$  \cite{Alkalaev:2016fok}. Note that just sending the central charge to infinity while keeping the dimensions fixed yields  the so called {\it light} blocks which are generically different from the global blocks by the truncated Virasoro character prefactor (see \cite{Alkalaev:2016fok} and \cite{Cho:2017oxl} for details, and our discussion below \eqref{laurent}). In what follows we calculate  the global blocks from scratch using the definitions \eqref{necklace} and \eqref{stickman} and restricting the Virasoro generators to $sl(2)$ ones. \footnote{On the other hand, global torus blocks can be viewed as solutions to the second order Casimir equations  arising from the $sl(2)$ Ward identities on a torus \cite{Kraus:2017ezw}. See also related considerations of global blocks within the recursive representations \cite{Cho:2017oxl}}  

\paragraph{$s$-channel.} Associating the 2-point $s$-channel block \eqref{necklace} to the $sl(2)$ algebra  we find 
\be
\label{globs}
\cF^{\Delta_{1,2}, \tilde \Delta_{1,2}}(q, z_{1,2}) = \sum_{n,m=0}^\infty \frac{\tau_{n,m}(\tilde\Delta_1, \Delta_1, \tilde\Delta_2)\tau_{m,n}(\tilde\Delta_2, \Delta_2, \tilde \Delta_1)}{n!\,m!\,(2\tilde\Delta_1)_n (2\tilde\Delta_2)_m}\, q^n\, x^{n-m}\;,
\ee
where $x=z_1/z_2$ and  the coefficients $\tau_{m,n} = \tau_{m,n}(\Delta_a, \Delta_b, \Delta_c)$ defining  the $sl(2)$ 3-point function of  a primary operator $\Delta_b$ and  descendant operators  $\Delta_{a,c}$ on the levels $n,m$ are given by \cite{Alkalaev:2015fbw}
\be
\label{Atau}
\tau_{n,m}(\Delta_{a,b,c}) = \sum_{p = 0}^{\min[n,m]} \frac{n!}{p!(n-p)!} (2\Delta_c +m-1)^{(p)} m^{(p)}
(\Delta_c+\Delta_b - \Delta_a)_{m-p}(\Delta_a + \Delta_b -\Delta_c+p-m)_{n-p}\;,
\ee
where $(a)_k = a(a+1)...(a+k-1)$ and $(a)^{(k)} = a(a-1)... (a-k+1)$. Setting $\Delta_2 = 0$, $\Delta_1 \equiv \Delta$ and  equating $\tilde\Delta_1 = \tilde\Delta_2 \equiv \tilde \Delta$ we will reproduce the 1-point torus block. Indeed, in this case $\tau_{n,m}(\tilde \Delta, 0, \tilde \Delta)  = \delta_{n,m} m!(2\tilde\Delta)_m $ and therefore the $n$-th global block coefficient \eqref{globs} is given by $\frac{\tau_{n,n}(\tilde\Delta, \Delta, \tilde\Delta)}{n!(2\tilde\Delta)_n}$. It defines the 1-point global block with external/exchanged dimensions $\Delta$ and $\tilde\Delta$ and can be expressed as the hypergeometric function coefficients \cite{Hadasz:2009db}. 

In the $n$-point case we can consider the generalized $s$-channel block defined by a diagram consisting of a loop with $n$ external legs. Let  $x_i = z_{i}/z_{i-1}$, where  $x_i = 2,...,n$ and $q = x_1 x_2 \cdots x_n$. With the identification $ \Delta_1 = \Delta_{n+1}$, $\tilde \Delta_1 = \tilde\Delta_{n+1}$ and $s_1 = s_{n+1}$ the $n$-point global block in the generalized $s$-channel is given by   
\be
\label{nglobs}
\cF^{\Delta_{i}, \tilde \Delta_{j}}(x) = \sum_{s_1,...,s_{n-1}=0}^\infty \;\;\prod_{m=1}^n \frac{\tau_{s_m,s_{m+1}}(\tilde\Delta_{m}, \Delta_m, \tilde\Delta_{m+1})}{s_m!\,(2\tilde\Delta_m)_{s_m} }  x_m^{s_m}\;,
\ee
where $\tau$-coefficients are given by \eqref{Atau}.

\paragraph{$t$-channel.} To find the 2-point global block in the $t$-channel \eqref{stickman} we use the OPE for quasi-primary fields, see e.g.  \cite{Belavin:1984vu,Blumenhagen,Qualls:2015qjb}, 
\be
\label{OPEsl}
\phi_1(z_1)\phi_2(z_2) = \sum_{\tilde\Delta_2} C_{_{\Delta_1\Delta_2\tilde\Delta_2}} (z_1-z_2)^{\tilde\Delta_2 - \Delta_1-\Delta_2}\, {}_1F_{1}\left(\tilde\Delta_2 + \Delta_1-\Delta_2,\, 2\tilde\Delta_2\big|(z_1-z_2)L_{-1}\right)\phi_{\tilde\Delta_2}(z_2)\;,
\ee 
where the OPE coefficients are now packed into the confluent hypergeometric function, cf. \eqref{OPE}--\eqref{OPEpsi}. Substituting the above OPE into the block function  \eqref{stickman} restricted to the $sl(2)$ subalgebra  we find the global $t$-channel block  
\be
\label{globt}
\cF^{\Delta_{1,2}, \tilde \Delta_{1,2}}(q, z_{1,2}) =\sum_{n,m=0}^\infty \frac{\sigma_{m}(\Delta_{1}, \Delta_{2}, \tilde\Delta_2)\tau_{n,n}(\tilde\Delta_1, \tilde\Delta_2, \tilde\Delta_1)}{n!m!(2\tilde\Delta_1)_n(2\tilde\Delta_2)_m} \, q^n \, w^m\;,
\ee
where $w = (z_1-z_2)/z_2$,  the $\tau$-function is given by \eqref{Atau}, the $\sigma$-function reads
\be
\sigma_{m}(\Delta_{1}, \Delta_{2}, \tilde\Delta_2) = (-)^m (\tilde\Delta_2+\Delta_1-\Delta_2)_{m}(\tilde \Delta_2)_{m} \;.
\ee
Setting $\Delta_1 = 0$, $\Delta_2  = \tilde \Delta_2 \equiv \Delta$, and $\tilde\Delta_1 \equiv \tilde \Delta$ we will reproduce the 1-point torus block because in this case $\sigma_{m}(0, \Delta,  \Delta)  = \delta_{m,0}$ and, therefore, the $n$-th global block coefficient \eqref{globt} is given by $\frac{\tau_{n,n}(\tilde\Delta, \Delta, \tilde\Delta)}{n!(2\tilde\Delta)_n}$ which is exactly the 1-point block coefficient. Let us note that the 1-point torus block factorizes from the 2-point expression. Moreover, \eqref{globt} is a product of two hypergeometric functions in accordance with the analysis in \cite{Kraus:2017ezw}. The higher-point generalizations can be obtained in the standard fashion by successively applying the OPE \eqref{OPEsl} in the torus  correlation function  \eqref{npt_dec} associated to the $sl(2)$ algebra.

\section{Exponentiating global blocks}
\label{sec:exp}
Let us consider the  regime of large dimensions, when all conformal dimensions are rescaled using a large parameter $\kappa$ in a coherent manner,
\be
\Delta_i = \varkappa\, \sigma_i\;, \qquad \tilde \Delta_i = \varkappa\, \tilde \sigma_i\;,\quad\qquad \varkappa \gg 1\;,
\ee
where  $\sigma_i$ and $\tilde \sigma_j$ can be referred to as {\it  classical global} dimensions. We expect that in this regime the global blocks are exponentiated
\be
\label{expoglob}
\cF^{\Delta_{i}, \tilde \Delta_{j}}(x) \, \cong\,  \exp\big[ \varkappa\, g^{\sigma_i, \tilde \sigma_j}(x)\big]\qquad  \text{as}\qquad  \varkappa \to \infty\;,
\ee
where $g^{\sigma_i, \tilde \sigma_j}(x)$ is a {\it  classical global} block. Indeed, $n$-point global blocks satisfy the Casimir channel equations which are second order partial differential equations with coefficients being rational functions of the conformal dimensions $\Delta_i, \Delta_j$ \cite{Dolan:2011dv, Alkalaev:2015fbw,Alkalaev:2016fok,Kraus:2017ezw}. At the same time from the general theory of differential equations it follows that once the equation coefficients depend on some large parameter in a specific way then  the leading asymptotics in the solution space are given by exponentials. \footnote{See e.g. the monograph \cite{Cesari}. The  1-point case was considered in \cite{Alkalaev:2016fok},  the $n$-point case  was discussed in \cite{Kraus:2017ezw}.} 

Matching  the central charge $c$ of the Virasoro block and the scale $\kappa$ of the global blocks we see that to some extent the classical global blocks \eqref{expoglob} are similar to the standard classical blocks, cf. \eqref{ccb}.\footnote{Contrary to the case of global blocks,  the Virasoro blocks do not satisfy any differential equations (except for special values of conformal dimensions that results in the BPZ equation). In particular, this is why the exponentiation of the Virasoro blocks has not been rigorously proven yet (for related considerations see \cite{Litvinov:2013sxa,Teschner:2017rve}). } Most probably, the standard classical block and classical global blocks are not related  for general values of conformal dimensions. However, there is a lot of evidence that they can be related to each other in particular perturbative regimes when some conformal dimensions $\Delta_{bgr}$ are much larger than the others $\Delta_{prt}$, 
\be
\label{ass}
\Delta_{prt}/\Delta_{bgr}\ll 1\;.
\ee
(Note that this assumption is equally translated both to the standard classical and global classical dimensions.) For example, in the sphere case with two background external operators the perturbative classical block coincides with respective perturbative classical global block \cite{Fitzpatrick:2015zha,Alkalaev:2015fbw}. In the $1$-point torus case, the relevant perturbation theory assumes that the exchanged channel (loop) dimension is much large than the external dimension. In this case we also observe that the perturbative classical block can be reproduced from the global block  \cite{Alkalaev:2016fok}.

\paragraph{Superlight expansion.} Using the superlight expansion assumption \eqref{constraints1} rewritten in terms of classical global dimensions $\sigma_i$ and $\tilde\sigma_j$ we find out that modulo the logarithmic term the respective perturbative classical block coincides with the perturbative classical global block. Indeed, the corresponding classical global block is given by  
\be
\label{g1}
g^{\tilde\sigma,\delta,\nu}(q, x) = \tilde \sigma \sum_{n=1}^\infty f^{(1)}_{2n}(x,q) \delta^{2n} +  \tilde\sigma \nu \sum_{n=2}^\infty f^{(2)}_n(x,q)\delta^n + \cO(\tilde\sigma^2, \nu^2)\;,
\ee
where the first term defines  the 1-point block with coefficients \eqref{f2f4},
and the second term is the first order $\nu$-correction, where a few  first coefficients read
\be
f^{(2)}_2(x,q) = \frac{x}{2}+\frac{1}{2}\sum_{n=1}^\infty \left[x+\frac{1}{x}\right] q^n \;,
\quad
f^{(2)}_3(x,q) = \frac{x^2}{8}+\frac{1}{8}\sum_{n=1}^\infty \left[(n+1)x^2 - 2n + \frac{n-1}{x^2}\right] q^n\;,
\ee
\be
f^{(2)}_4(x,q) = \frac{x^3}{24}+\frac{1}{48}\sum_{n=1}^\infty\left[(n+1)(n+2) x^3  - 3n(n+1)x - \frac{3(n-1)n}{x} + \frac{(n-1)(n-2)}{x^3}\right]q^n \;.
\ee
We observe  that in a given order this expression coincides with  the perturbative conformal block calculated  using  the combinatorial representation \eqref{bloook}.  Also, as a consistency check,  setting $x=1$ we reproduce the 1-point perturbative torus block with the external conformal dimension $(\epsilon_1+\epsilon_2)$. Technically, at this point the second term in \eqref{g1} coincides with the first term up to a prefactor fixed in the linear order in $\nu$ by that $(\frac{\epsilon_1+\epsilon_2}{\tilde\epsilon})^n\approx (1+ n  \nu)\delta^n$.   Indeed, we notice that the polynomial in the square bracket of $f^{(2)}_3$ has a root  $x=1$, and, therefore,  we find that $f^{(2)}_{2k+1}(x=1) = const$ and  $f^{(2)}_{2k}(x=1) =   2k f^{(1)}_{2k}+const$. 

\vspace{-1mm}

\paragraph{Double leg expansion.} Similarly, using the double leg expansion assumption \eqref{constraints2} rewritten in terms of the classical global dimensions $\sigma_i$ and $\tilde\sigma_j$ we find out that  the corresponding block is given by    
\be
\label{g2}
g^{\tilde\sigma,\delta}(q, x)  = \tilde \sigma \sum_{n=1}^\infty g_{n}(x,q) \delta^{n} + \cO(\tilde\sigma^2)\;,
\ee
where a few first coefficients are found to be 
\be
g_{2}(x,q) =\frac{x}{2}+ \frac{1}{2} \sum_{n=1}^\infty\left[x+\frac{1}{x}+2\right]q^n\;,
\quad
g_{3}(x,q)=\frac{x^2}{4}+ \frac{1}{4} \sum_{n=1}^\infty\left[(n+1)x^2-2n +\frac{n-1}{x^2}\right]q^n\;,
\ee
\be
\ba{l}
\dps
g_{4}(x,q)= \frac{10 x^3-3 x^2}{48} +\frac{1}{48}\sum_{n=1}^\infty \bigg[5(n+1)(n+2)x^3-3(n+1)^2 x^2 -9n(n+1)x
\\
\\
\dps
\hspace{40mm} -2(n-1)(n+1) - \frac{9n(n-1)}{x}-\frac{3(n-1)^2}{x^2}+\frac{5(n-1)(n-2)}{x^3}\bigg]q^n\;.
\ea
\ee
In a given order, the above expression reproduces  the perturbative conformal block calculated using the combinatorial representation \eqref{classblockhigh}. Noticing that $g_3 = 2f^{(2)}_3$ and setting $x=1$ we are left with  the 1-point perturbative torus block with the external conformal dimension $2\epsilon$ because $g^{(1)}_{2k+1}(x=1) = const$ and $g^{(1)}_{2k}(x=1) =  2^{2k} f^{(1)}_{2k}+const$. Indeed, going to the 1-point case produces factors $(\frac{2\epsilon}{\tilde\epsilon})^n\approx 2^n \delta^n$. 

\vspace{-1mm}

\paragraph{t-channel case.} The relevant perturbation theory in the $t$-cannel leads to imposing the following constrains,
\be
\label{tfusion}
\epsilon_1 = \epsilon_2\;,
\qquad
\delta = \frac{\tilde\epsilon_2}{\epsilon_1} \ll1\;,
\ee
where the first constraint is the fusion rule guaranteeing that  $\tilde \epsilon_2=0$ approximation is  consistent. The corresponding perturbative classical global block is found to be  
\be
g^{\tilde\sigma_{1}, \delta}(q,w) = \tilde \sigma \sum_{n=1}^\infty  h_n(q,w) \delta^n+ \cO(\tilde\sigma^2)\;,
\ee 
where a first few coefficients can be represented as 
\be
h_1(q,w)=\text{Arccosh}\frac{i}{\sqrt{w}}-\log \frac{2 i}{\sqrt{w}}\;, 
%= \frac{  w}{2}-\frac{3   w^2}{16}+\frac{5  w^3}{48}-\frac{35   w^4}{512}+ \cO(w^5)\;,
\ee
\be
h_2(q,w) = \frac{q}{2}\frac{1}{1-q}\;,
\qquad
h_3(q,w) = -\frac{q^2}{48}\frac{q-3}{(1-q)^3}\;,\qquad ...\;.
\ee
We notice that the first coefficient $h_1$ does not depend on $q$, while higher coefficients $h_n$ at $n\geq 2$ do not depend on $w$ and coincide with the 1-point coefficients \eqref{f2f4}. 

\section{Holographic duals of the classical $s$-channel torus blocks}
\label{sec:dual}

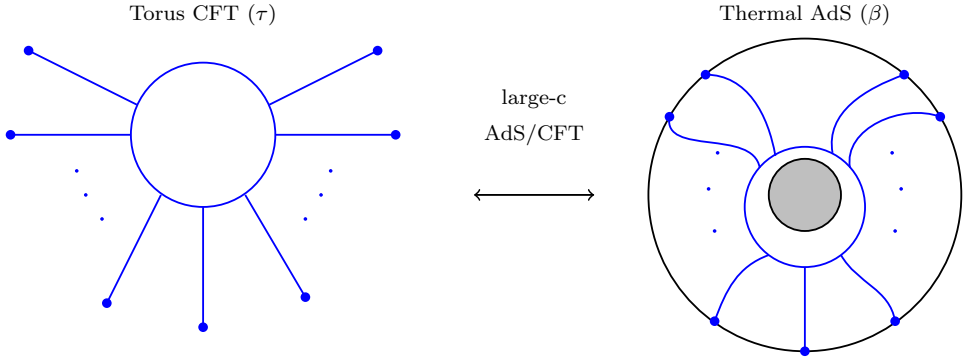
\begin{figure}[H]
\centering

\begin{tikzpicture}[line width=0.7pt,scale=0.80]

%%% block %%%%

\draw[blue] (-11,0) circle (1.2cm);
%\draw[blue] (-9,0) circle (3.2cm);
%\draw[blue] (-9,0) circle (2.2cm);

\draw[blue] (-11,-1.2) -- (-11,-3.2);
\fill[blue] (-9.3,-2.7) circle (0.8mm);

\draw[blue] (-10.3,-1.0) -- (-9.3,-2.7);
\fill[blue] (-11,-3.2) circle (0.8mm);

\draw[blue] (-11.7,-1.0) -- (-12.6,-2.8);
\fill[blue] (-12.6,-2.8) circle (0.8mm);

\draw[blue] (-14.2,0) -- (-12.2,0);
\fill[blue] (-14.2,0) circle (0.8mm);

\fill[blue] (-7.8,0) circle (0.8mm);
\draw[blue] (-9.8,0) -- (-7.8,0);

\fill[blue] (-8.1,1.4) circle (0.8mm);
\draw[blue] (-8.1,1.4) -- (-9.9,0.5);

\fill[blue] (-13.9,1.4) circle (0.8mm);
\draw[blue] (-13.9,1.4) -- (-12.1,0.5);

\fill[blue] (-13.1,-0.6) circle (0.3mm);
\fill[blue] (-12.95,-1.0) circle (0.3mm);
\fill[blue] (-12.68,-1.4) circle (0.3mm);

\fill[blue] (-8.88,-0.6) circle (0.3mm);
\fill[blue] (-9.04,-1.0) circle (0.3mm);
\fill[blue] (-9.3,-1.4) circle (0.3mm);

%\draw (-10.7,0) node {$\tilde\Delta$};
%\draw (-8.4,-2.3) node {$\Delta$};
%\draw (-10,-3.6) node {$z=1$};

\draw (-11,2) node {$\text{\scriptsize Torus CFT ($\tau$)}$};

%%%% arrow %%%%%

\draw (-5.5,0.6) node {\text{\scriptsize large-c}};
\draw (-5.5,0) node {$\text{\scriptsize AdS/CFT}$};

\draw[<->] (-6.5,-1.0) -- (-4.5,-1.0);

%\draw (-5,-1.5) node {$\hat\tau  \leftrightarrow \tau $};

%%%% annulus %%%%

\draw (-1,2) node {$\text{\scriptsize Thermal AdS ($\beta$)}$};

%\draw (0,2) node {$\text{Thermal AdS}\;\; (\,\tau\,)$};

\draw[fill=lightgray] (-1,-1.0) circle (0.6cm);
\draw (-1,-1.0) circle (2.6cm);
%\draw (-1,-1.0) circle (1.6cm);

\draw[blue] (-1,-1.2) circle (1.0cm);
\draw[blue] (-1,-2.2) -- (-1,-3.6);
\fill[blue] (-1,-3.6) circle (0.8mm);

\draw[blue] (-2.5,-3.1) to [out=70,in=200] (-1.6,-2.0);
\fill[blue] (-2.5,-3.1) circle (0.8mm);

\draw[blue] (0.5,-3.1) to [out=100,in=300] (-0.4,-2.0);
\fill[blue] (0.5,-3.1) circle (0.8mm);

\draw[blue] (-0.54,-0.3) to [out=100,in=200] (0.65,1.0);
\fill[blue] (0.65,1.0) circle (0.8mm);

\draw[blue] (-1.49,-0.35) to [out=100,in=0] (-2.65,1.0);
\fill[blue] (-2.65,1.0) circle (0.8mm);

\draw[blue] (-3.25,0.3) to [out=-100,in=100] (-1.75,-0.55);
\fill[blue] (-3.25,0.3) circle (0.8mm);

\draw[blue] (1.25,0.3) to [out=-200,in=100] (-0.25,-0.55);
\fill[blue] (1.25,0.3) circle (0.8mm);

\fill[blue] (0.45,-0.3) circle (0.3mm);
\fill[blue] (0.6,-0.9) circle (0.3mm);
\fill[blue] (0.5,-1.6) circle (0.3mm);

\fill[blue] (-2.45,-0.3) circle (0.3mm);
\fill[blue] (-2.6,-0.9) circle (0.3mm);
\fill[blue] (-2.5,-1.6) circle (0.3mm);

\end{tikzpicture}
\caption{$n$-point $s$-channel conformal blocks with the modular parameter $\tau$ in the $s$-channel holographically realized as the necklace graph on two-dimensional slice of the thermal AdS. Time runs along the non-contractible cycle $t \sim t+ \beta$. Primary operators are inserted at $x_i$ at the boundary. These points are mapped to boundary attachment points $y_i$ in the bulk. } 
\label{duality}
\end{figure}

In this section we advocate that in the $s$-channel the perturbative classical block function $f_{class}$ considered in Section \bref{sec:heavy-light} can be represented as \footnote{The case of 1-point classical torus blocks was considered in \cite{Alkalaev:2016ptm}. The holographic correspondence for higher point global blocks was discussed recently in \cite{Kraus:2017ezw}.} 
\be
\label{identification}
-f_{class}(x,q)\, \cong\, S_{therm}+ L_{dual}(y,\beta)\;,
\ee 
where $S_{therm}$ is the holomorphic part of the 3d gravity action evaluated on the thermal AdS space,  $L_{dual}(y,\beta)$ is the length of the dual necklace graph attached to the  boundary points $y_i$, $i=1,...,n$, see Fig. \bref{duality}.  The gravitational action expressed in terms of the rescaled central charge $c/6$ is $S_{thermal} = i\pi \tau/2$ \cite{Carlip:1994gc,Maldacena:1998bw,Kraus:2006wn}. In terms of the modular parameter the action is $S_{thermal} = 1/4 \log q$ that reproduces the classical conformal block \eqref{class_block_1} at zeroth dimensions. The block/length correspondence \eqref{identification} is supplemented with the identification of the modular parameter and coordinates of the primary operators as 
\be
\label{map}
\beta = - \log q\;, 
\qquad
y_i = - \log x_i\;.
\ee

Let us consider the thermal AdS space that is a solid torus with  time running along the  non-contractible cycle,
\be
\label{thermal}
ds^2 = \big(1+ \frac{r^2}{l^2}\big)dt^2 +\big(1+ \frac{r^2}{l^2}\big)^{-1}dr^2 + r^2 d \varphi^2\;,
\ee
where    $t \sim t + \beta$, $\varphi \sim \varphi + 2\pi$, $r \geq 0$, the AdS radius is set $l=1$. The time period $\beta$ defines the modular parameter $\tau_{ads} = i \beta/2\pi$ and the temperature $\beta \sim T^{-1}$.  The equivalent form of the metric can be obtained by rescaling the time coordinate $t \rightarrow -i \tau_{ads} t$ so that the metric coefficients explicitly depend on the modular parameter while $t \sim t + 2\pi$. However,  the metric \eqref{thermal} is more convenient in practice because  the modular parameter shows up only when integrating along the  non-contractible cycle. Otherwise, the local  dynamics is the same as in the AdS space with non-periodic time.

Let us notice that in order to describe the correspondence   between holomorphic conformal blocks and geodesic networks on the two-dimensional slice \eqref{identification} we assume that the modular parameter $q$ and coordinates of the primary operators  are real \eqref{map}. In this way we obtain that the holomorphic conformal block, being by definition a complex function, is equated to the real geodesic length on the thermal AdS with pure imaginary modulus $\tau_{ads}$.\footnote{The geodesics network stretched between boundary points on different two-dimensional slices computes the full classical conformal block which is the sum of holomorphic and antiholomorphic conformal blocks. See \cite{Hijano:2015rla} for an extended discussion in the sphere case.} Since  $q = e^{2\pi i \tau_{cft}}$ we find that up to the modular transformations the parameter $\tau_{cft}$ can take one of two values   
$\tau_{cft} = 0+ i a$ or $\tau_{cft} = \half+ i a$, where $\forall a \in \mathbb{R}$.
Therefore, 
\be
\label{twoads}
\tau_{cft} = \tau_{ads} \qquad \text{or}\qquad  \tau_{cft} = \tau_{ads}+\half\;\;.
\ee 
In the first case, the torus where our CFT lives is indeed the conformal boundary of the bulk space. This is realized  by the map \eqref{map}. In the second case, the modular parameters are different but in the low-temperature approximation $\tau_{ads}\to \infty$, we again arrive at the standard duality with $\tau_{cft} \approx  \tau_{ads}$  \cite{Alkalaev:2016ptm}. The two corresponding  bulk solutions are discussed below in Section \bref{sec:711}.  

% On the other hand, we consider the thermal AdS with pure imaginary modulus $\tau_{ads}$. The analogous phenomenon is known in the sphere case where restricting to fixed time slice in the bulk restricts the coordinates of primary operators $\phi_i(z_i, \bar z_i)$ to lie on the circle $z_i\bar z_i  =1$ (cf. \eqref{map}, see also  Appendix B of \cite{Hijano:2015rla} for how to extend the correspondence for different times). }    

%Let us now shortly review the worldline formulation on the thermal AdS background with the metric \eqref{thermal}, see  \cite{Hijano:2015rla,Alkalaev:2015wia,Alkalaev:2016ptm} for more details. A geodesic segment can  be described by the   action 
%\be
%\label{total2}
%S = \int_{\lambda_1}^{\lambda_2} d\lambda \sqrt{g_{mn}(x)\dot x^m \dot x^n}\;,
%\ee
%where local coordinates are $x^m = (t,\phi,r)$ and the metric coefficients $g_{mn}(x)$ are read off from \eqref{thermal}. The reparametrization invariance allows us to impose the normalization condition $|g_{mn}(x)\dot x^m \dot x^n| = 1$ so that the on-shell action is now given by $S = \lambda_2 - \lambda_1$. Using the Killing vectors of \eqref{thermal} we can restrict the dynamics to the constant angle $\varphi =0$ surface (the annulus on Fig. \bref{duality}). The corresponding conserved momentum is $p_\phi = 0$, while the other conserved momentum $p_t$ denoted below as $s = |p_t|$ is the integration  constant that defines  the shape of geodesics. See Appendix \bref{sec:worldline} for more details.   

\subsection{Worldline formulation}
\label{sec:worldline}

In what follows we shortly review the worldline formulation on the thermal AdS background with the metric \eqref{thermal}, see  \cite{Hijano:2015rla,Alkalaev:2015wia,Alkalaev:2016ptm} for more details. A geodesic segment can  be described by the   action 
\be
\label{total2}
S = \int_{\lambda_1}^{\lambda_2} d\lambda \sqrt{g_{mn}(x)\dot x^m \dot x^n}\;,
\ee
where local coordinates are $x^m = (t,\phi,r)$ and the metric coefficients $g_{mn}(x)$ are read off from \eqref{thermal}. The reparametrization invariance allows us to impose the normalization condition $|g_{mn}(x)\dot x^m \dot x^n| = 1$ so that the on-shell action is now given by $S = \lambda_2 - \lambda_1$. Using the Killing vectors of \eqref{thermal} we can restrict the dynamics to the constant angle $\varphi =0$ surface (the annulus on Fig. \bref{duality}). The corresponding conserved momentum is $p_\phi = 0$, while the other conserved momentum $p_t$ is the motion constant that defines  the shape of geodesics. In this case, the normalization condition is given by  
\be
\label{dotr}
\dot r  =  \sqrt{r^2 -s^2+1}\;,\qquad\;\; s \equiv |p_t|\;.
\ee

\begin{itemize}

\item The length of a loop segment is given by    
\be
\label{sloop}
L_{loop} = \frac{1}{\tilde s}\int^{t_2}_{t_1} dt \big(1+r^2(t)\big)\;,
\ee
where $r(t)$ being the radial deviation, and $t_{1,2}$ are initial/final time positions. 

\item The length of an external leg with one endpoint attached to the conformal boundary is
\be
\label{leg}
L_{leg} = \int_{\rho}^{\Lambda} \frac{dr}{\sqrt{1+r^2-s^2}} = -\log(\rho+ \sqrt{\rho^2 -s^2+1})\;,
\ee
where $\rho$ is the vertex radial coordinate, and the cutoff parameter $\Lambda\to \infty$ is introduced to regularize the conformal boundary position. 

\end{itemize}

\noindent The radial deviation is governed by the following evolution equation  
\be
\label{diff}
\frac{dr}{dt} = \pm\,\frac{1}{ s}(1+r^2)\sqrt{r^2 -  s^2+1}\;,
\ee 
that can be explicitly integrated. Using \eqref{diff} we can express the momentum $s$ in terms of the radius and its time derivative  and substitute then into \eqref{sloop}. The resulting action is  \eqref{total2} on the $\phi = const$  slice, where the metric is given by \eqref{thermal}. The corresponding equations of motion are second-order ODE having a general solution that depends on two integration constants. The formulation described above is partially integrated form of these equations. Indeed, one of integration constants is given by the time momentum that drastically simplifies the analysts of geodesic lines, while the residual dynamical equation \eqref{diff} is first-order ODE with a general solution depending on the other integration constant.    

Integrating the evolution equation \eqref{diff} we get  
\be
\label{angsep}
e^{2 (t-t_0)}=\frac{\left(i+r_0\right)(r(t)-i) \left(r_0 -s \sqrt{r_0^2-s^2+1}+is^2-i\right) \left(s \sqrt{r(t)^2-s^2+1}+r(t)-i s^2+i\right)}{(i-r_0) (r(t)+i) \left(r_0 +s \sqrt{r_0^2-s^2+1}-i s^2+i\right) \left(s \sqrt{r(t)^2-s^2+1}-r(t)-is^2+i\right)}\;,
\ee
where $r(t_0) = r_0$ are initial conditions. Solving this equation we can express the radial coordinate as a function of parameters $r_0$ and $s$, i.e. $r = r(t|r_0,s)$. 

\subsection{Dual geodesic networks}

Let us  first describe the kinematics of the dual networks in the $n$-point case. The  graph on Fig. \bref{duality} is drawn on two-dimensional annulus with coordinates $(r, t)$ and  consists of $n$ external legs and $n$ loop segments stretched between $n$ vertex points. Any trivalent vertex connects two loop segments and one external leg. The radial and time vertex coordinates are denoted as $(\rho_m, \gamma_m)$,  $m = 1,...,n$. The external legs are attached to the conformal boundary at points $y_m$, where $m =1,...,n$. The motion constants (time momenta) of  loop segments and external legs are respectively $\tilde s_m$ and $s_k$, where $m,k = 1, ..., n$.

The geodesic length can be associated to the mechanical system of massive test particles propagating in the bulk space. Hence,  the total length of the geodesic network is given by  
\be
\label{tot0}
L_{dual} = \sum_{m=1}^n \left(\,\epsilon_m\, L_{leg}^m+ \tilde \epsilon_m\, L^m_{loop}\,\right)\;,
\ee
where  classical conformal dimensions $\sim$ masses, and  the length of each geodesic segment \eqref{sloop}, \eqref{leg} reads 
\be
\label{length_segments}
L^m_{loop} =  \frac{1}{\tilde s_m}\int_{\gamma_1^m}^{\gamma_2^m} dt \left(1+R_m^2(t)\right)\;, 
\qquad
L_{leg}^m = -\log(\rho_m+\sqrt{\rho_m^2-s_m^2+1})\;, 
\ee 
where $R_m(t)$ is a radius of $m$-th loop segment, $s_m$ and $\tilde s_m$  are the leg and loop momenta, $\rho_m$ is the radial vertex coordinate, the integration limits $\gamma_{1,2}^m$ parameterize endpoints of each loop segment: 
\be
\label{rules2}
\gamma_1^m  \equiv \gamma_m\;, 
\quad 
\gamma_2^m\equiv \gamma_{m+1}\;, \qquad  m=1,..., n\;,
\ee
\be
\label{rules}
\gamma_{n+1}\equiv \gamma_1+\beta\;.
\ee 
To find the total length of the geodesic network we have to  know  the radial functions $R_m(t)$ explicitly as well as  the vertex position $\rho_m$, the momenta $s_m, \tilde s_m$, and the times $\gamma_m$.

\paragraph{Time evolution.} The radial functions $R_m(t)$ in \eqref{length_segments} are subjected to the evolution equations \eqref{diff},
\be
\label{evoleq}
\frac{d R_m}{dt} = \frac{1}{ \tilde s_m} (1+R_m^2)\sqrt{R^2 - \tilde s_m^2+1}\;,\qquad m=1,...,n\;,
\ee
whose  general solution $R_m = R_m(t\,|\, c_m, \tilde s_m)$ is parameterized by two constants, an integration constant $c_m$ and the loop momentum $\tilde s_m$. In order to fix them we impose the vertex   boundary conditions   
\be
\label{boundar0}
R_m(\gamma_m) = \rho_m\;, 
\qquad
R_{m}(\gamma_{m+1}) = \rho_{m+1}\;, 
\qquad 
m = 1,...,n\;,
\ee
supplemented with the condition  \eqref{rules}. Whence, we have $2m$ equations for $2m$ variables that allows us to solve the evolution equations in terms of the vertex radial positions.  

\paragraph{Time intervals.} In order to find  initial/final time positions \eqref{rules2}--\eqref{rules} explicitly we use the general formula for time intervals stretched between given two radial positions \eqref{angsep}. Let us consider the $m$-th leg stretched from the vertex point $(\rho_m, \gamma_m)$ to the boundary attachment point $(\infty, y_m)$. In this case, using  formula \eqref{angsep} we find the time interval relation    
\be
\label{angvert}
e^{2(\gamma_m-y_m)} =\frac{\rho^2_m+\left(\rho^2_m-1\right) s_m^2-2 \rho_m s_m \sqrt{\rho_m^2-s_m^2+1}+1}{\left(\rho_m^2+1\right) (s_m-1)^2}\;,
\qquad
m=1,...,n\;,
\ee    
which can be explicitly solved as $\gamma_m = \gamma_m(\rho_m, s_m, y_m)$.

\paragraph{Momentum conservation conditions.} Both the vertex radial positions and momenta are constrained by the conservation conditions following from the least action principle for the total action \eqref{tot0}. There are three possible configurations of external legs according to values of their momenta: radial, convex, or concave legs. In what follows we consider the case of one concave and $n-1$ convex legs. Other possible options can be shown to be equivalent to this configuration or be inconsistent. 

The momentum conservation condition in each vertex reads $\tilde \epsilon_m \tilde p_m + \tilde \epsilon_{m+1} \tilde p_{m+1} + \epsilon_m p_m =0$, where $m=1,...,n$ and $p_m, \tilde p_m$ are two-component conserved  momenta on the annulus (for the general discussion of trivalent vertices on the hyperbolic spaces see \cite{Alkalaev:2016rjl}). The corresponding  time and radial components are given by  
\be
\label{consn1}
\epsilon_m s_m+\tilde \epsilon_m \tilde s_m -\tilde \epsilon_{m+1}\tilde s_{m+1} =0\;,
\ee 
\be
\label{consn2}
\tilde\epsilon_m \sqrt{\rho_m^2-\tilde s_m^2+1} +\tilde\epsilon_{m+1} \sqrt{\rho_m^2-\tilde s_{m+1}^2+1}- \epsilon_m \sqrt{\rho_m^2- s_m^2+1}=0\;,
\ee
where indices run $m = 1,...,n$ with the  identification $(n+1)\to 1$. Remarkably, these constraints can be explicitly solved as
\be
\label{general}
s_ m  = \frac{\tilde \epsilon_m\tilde s_m - \tilde \epsilon_{m+1}\tilde s_{m+1}}{\epsilon_m}\;,
\qquad
\rho^2_m = -\frac{(\tilde s_m^2+\tilde s_{m+1}^2-2 \sigma_m \tilde s_m \tilde s_{m+1})}{1-\sigma_m^2}-1\;, 
\ee
where we introduced 
\be
\sigma_m  = \frac{\tilde\epsilon_m^2 + \tilde\epsilon_{m+1}^2 - \epsilon_m^2}{2\tilde\epsilon_{m}\tilde\epsilon_{m+1}}\;.
\ee
Following the general analysis of trivalent vertices on hyperbolic spaces \cite{Alkalaev:2016rjl} we can show that there are real vertex solutions $\rho_m^2\geq 0 $ provided that the conformal dimensions satisfy the triangle inequalities
\be
\tilde\epsilon_m+ \tilde\epsilon_{m+1}\geq \epsilon_m\;,
\qquad
\tilde\epsilon_m+ \epsilon_{m}\geq \tilde \epsilon_{m+1}\;,
\qquad
\tilde\epsilon_{m+1}+ \epsilon_{m}\geq \tilde \epsilon_m\;.
\ee 

\paragraph{Final assembling.} Now, we gather all ingredients discussed above. In general, there are $2n$ equations \eqref{boundar0}, $2n$ equations \eqref{consn1} and \eqref{consn2}, and  $n$ equations \eqref{angvert} for $5n$ variables $\rho_m$, $s_m, \tilde s_m$, $\gamma_m$, and integration constants $c_m$, $m=1,...,n$. Thus, we unambiguously fix all these variables in terms of the modular parameter $\beta$, the boundary attachment positions $y_m$, and classical dimensions $\epsilon_m, \tilde \epsilon_m$, $m=1,...,n$. For convenience, using the rotational invariance of the boundary circle we can set $y_1=0$. Finally, we will obtain the total length as $L_{dual} = L_{dual}(\beta, y_m, \epsilon_m, \tilde\epsilon_m)$.

\section{Dual networks for the 2-point block}
\label{sec:sat}

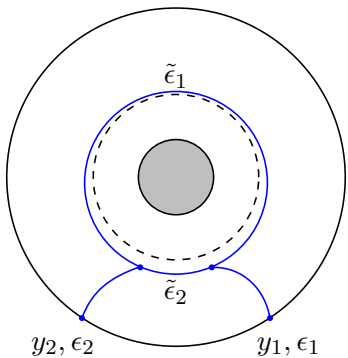
\begin{figure}[H]
\centering
\begin{tikzpicture}[line width=0.6pt, scale = 0.50]
\draw (0,0) circle (4.5cm);
\draw[fill=lightgray] (0,0) circle (1cm);

\draw[dashed] (0,0) circle (2.2cm);
\draw[blue] (0,-0.15) circle (2.43cm);

%%%% first line

%\fill[blue] (-1.85,-4.1) circle (0.8mm);
\fill[blue] (-0.95,-2.4) circle (0.8mm);

%\draw[blue] (-1.85,-4.1) --  (-1.0,-2.4);
\draw[blue] (-2.5,-3.75) to [out=70,in=200] (-1.0,-2.4);
\fill[blue] (-2.5,-3.75) circle (0.8mm);

%%%% second line

\fill[blue] (2.5,-3.75) circle (0.8mm);
\fill[blue] (0.95,-2.4) circle (0.8mm);

%\draw[blue] (1.85,-4.1) -- (1.0,-2.4);
\draw[blue] (2.5,-3.75) to [out=100,in=0] (1.0,-2.4);

\draw (0.0,2.7) node {$\tilde \epsilon_1$};
\draw (-0.0,-3.05) node {$\tilde\epsilon_2$};
\draw (-3,-4.5) node {$y_2,\epsilon_2$};
\draw (3,-4.5) node {$y_1,\epsilon_1$};

\end{tikzpicture}
\caption{The necklace graph on the annulus. The inner and outer black solid circles represent the conformal boundary. The  blue dashed circle goes along the $r=0$ radius and corresponds to the 0-point block. The blue solid loop is a deformation of the dashed circle by external operators. On the conformal boundary, we choose $y_1=0$ and denote $y_2\equiv y$.} 
\label{annulus}
\end{figure}

We shall apply the general scheme described above to the 2-point case. The respective  geodesic network is shown on Fig. \bref{annulus}. It consists of four different segments: two loop segments and two legs stretched from the boundary attachment points to trivalent vertex points.  
\noindent  The total length is given by  
\be
\label{tot}
L_{dual} = \tilde \epsilon_1 L^1_{loop}+ \tilde \epsilon_2 L^2_{loop} + \epsilon_1 L_{leg}^1+ \epsilon_2 L_{leg}^2\;,
\ee
where the segments are described by \eqref{length_segments}--\eqref{rules}. The radial coordinates of  each loop segment $R_m(t)$ in \eqref{length_segments} are described by the evolution equation \eqref{evoleq} subjected to the vertex boundary conditions \eqref{boundar0}, which in this case take the form 
\be
\label{boundar}
R_1(\gamma_1)=\rho_1\;,
\qquad
R_1(\gamma_2) = \rho_2\;,
\qquad\;\;\;\; 
R_2(\gamma_2)=\rho_2\;,
\qquad
R_2(\gamma_1+\beta) = \rho_1\;.
\ee
The general solutions $R_1 = R_1(t\,|\, c_1, \tilde s_1)$ and $R_2 = R_2(t\,|\, c_2, \tilde s_2)$ 
are parameterized by two integrations constants $c_1$ and $c_2$. 

On top of that there are four momentum conservation conditions \eqref{consn1} and \eqref{consn2}. Thus, we have to solve 8 equations  to fix 8 variables $\rho_{1,2}$, $s_{1,2}$, $\tilde s_{1,2}$, and $c_{1,2}$. Exact solutions to this algebraic systems are not known that is a common problem arising  in the bulk analysis, for a review see, e.g. \cite{Alkalaev:2016rjl}. 

In what follows we use different approximations analogous to those  developed  for the classical conformal blocks in Section \bref{sec:heavy-light}. In general, we assume that the loop segments are more massive than the external legs, 
\be
\tilde \epsilon_m/\epsilon_k\ll 1\;.
\ee  
To simplify our consideration we  assume that  $\tilde\epsilon_k = \tilde\epsilon_m\equiv \tilde \epsilon$ that gives rise to the natural heavy parameter. Expanding around $\tilde \epsilon = \infty$ we arrive at the perturbative geodesic network length function which is linear in $\tilde\epsilon$ and depends on  $\delta_i = \epsilon_i/\tilde\epsilon\ll 1$. In the 2-point case, there are  two small parameters that define further possible perturbation expansions: (i) $\delta_2 \ll \delta_1$, (ii) $\delta_1 = \delta_2$. In this section we explicitly consider these two opposite regimes referring to them respectively as superlight and double leg expansions. 

\subsection{Superlight expansion }

In this case, the 2-point block can be considered as a small perturbation of the 1-point block. Recalling the  constraints \eqref{constraints1},
\be
\label{epsilons}
\tilde\epsilon_1 = \tilde\epsilon_2\equiv \tilde \epsilon\;, \qquad \delta = \frac{\epsilon_1}{\tilde\epsilon} \ll 1\;, \qquad \nu = \frac{\epsilon_2}{\epsilon_1} \ll 1\;,
\ee
we expand any quantity $F$  as follows $F(\delta,\nu) = F_0(\delta) + \nu F_1(\delta) +\nu^2 F_2(\delta) + ... $, 
where $F_0(\delta)$ is a seed function corresponding to the 1-point case, while expansion coefficients  $F_i(\delta)$ are  expanded  in $\delta$.

\subsubsection{The seed solution: 1-point block}  
\label{sec:711}  
The seed solution arising at $\nu=0$ is given by the tadpole geodesic graph corresponding to the 1-point perturbative torus block \cite{Alkalaev:2016ptm}. However, it is not exact in $\delta$ because the remaining  geodesic equations are still too complicated to be solved exactly.

In the 1-point case the conservation  conditions are reduced to   
\be
\label{srho}
\tilde s_1 = \tilde s_2 \;, \qquad s_1 = 0\;, 
\qquad
\rho_1 =\sqrt{\frac{4\,\tilde s_1^2}{4-\delta^2}-1}\;.
\ee 
Following \cite{Alkalaev:2016ptm} we expand $\tilde s_1 \equiv s$, $R_1(t)\equiv R(t)$, and $\rho_1\equiv \rho$ as 
\be
\label{fin1}
\ba{l}
s = 1+ s_2 \delta^2 + s_4 \delta^4 + \cO(\delta^6)\;, 
\\
\\
R(t) = R_1(t) \delta^1 + R_3(t) \delta^3 + \cO(\delta^5)\;, 
\\
\\
\rho = \rho_1 \delta^1 + \rho_3 \delta^3 + \cO(\delta^5)\;. 
\ea
\ee 
We note that  the time coordinate of the vertex defined by \eqref{angvert} does not depend on $\delta$. Indeed,  $s_1 = 0$, i.e. the leg is stretched along the radial direction, and, therefore, from \eqref{angvert} it follows that $\gamma_2 = y_2 =0$. 

Now, we see that the boundary conditions \eqref{boundar0} in the 1-point case are given by  $R(0)=\rho$ and $R(0)=R(\beta)$. Using the first condition we find that the time interval equation \eqref{angsep} can be reduced to  
\be
\label{angsep1}
e^{2 t}=\frac{(-1+i \rho ) (R(t)-i) \left(-i \rho +i s \sqrt{\rho^2-s^2+1}+s^2-1\right) \left(s \sqrt{R(t)^2-s^2+1}+R(t)-i s^2+i\right)}{(\rho -i) (-1+i R(t)) \left(\rho +s \sqrt{\rho ^2-s^2+1}-i s^2+i\right) \left(i s \sqrt{R(t)^2-s^2+1}-i R(t)+s^2-1\right)}\;.
\ee
Expanding \eqref{srho}, \eqref{angsep1} in  $\delta$ using \eqref{fin1} we find in the first non-trivial order that there are {\it two} possible sets of solutions 
\be
\label{sol1}
\rho_1 = \frac{1}{2} \coth \frac{\beta}{2}\;,
\qquad
s_2 = \frac{1}{8} \text{csch}^2 \,\frac{\beta}{2}\;,
\ee 
\be
\label{sol2}
\;\;\rho_1 = \frac{1}{2} \tanh \frac{\beta}{2}\;,
\qquad
s_2 =  -\frac{1}{8} \text{sech}^2\,\frac{\beta}{2}\;,
\ee
which can be used to calculate  the corresponding  first order correction $R_1(t)$. 

The solution set \eqref{sol2} and the length of the corresponding tadpole graph has been analyzed in \cite{Alkalaev:2016ptm}. One can show that in this case the loop goes through the point $t = \beta/2$ and $r=0$ in all orders in $\delta$. It implies that the deformation caused by the  leg pulling the loop is rapidly damped near the vertex point. In what follows we use the solution set \eqref{sol1}. In this case the radial deviations of the  loop segments are not vanishing near $t=\beta/2$  meaning that the leg produces perturbations along the whole loop.\footnote{We are grateful to Per Kraus for useful discussion of this point.} Most importantly, these two types of solutions correspond to  boundary CFTs with two different modular parameters, see our discussion below \eqref{twoads}.    

To summarize,  the seed solution is given by  
\be
\label{seed1}
R(t)=\frac{1}{2}\left(\coth \frac{\beta }{2} \cosh t-\sinh t\right)\delta + \cO(\delta^3)
\qquad
\rho_{1} =\frac{\delta}{2}\coth\frac{\beta }{2} + \cO(\delta^3)\;,
\ee
\be
\label{seed2}
s_{1}=0\;,
\qquad
\tilde s_1 = \tilde s_2\;,
\qquad 
\tilde s_{1}  = 1+\frac{\delta^2}{8}  \text{csch}^2\,\frac{\beta }{2} + \cO(\delta^4)\;.
\ee
We note that the zeroth order   $s_{2}$, $\rho_{2}$, and $\gamma_{2}$ associated to the second vertex turn out to be non-vanishing power series in $\delta$, see below.\footnote{For consistency check, we can step aside from the first order formulation used to find \eqref{seed1}--\eqref{seed2} and reproduce the same answer solving the geodesic equations of motion that follow from the standard second-order formulation, see Section \bref{sec:worldline} for more details.} They can also be treated as the seed quantities  not seen at $\nu^0$ order.

\subsubsection{First order corrections}

Following Appendix \bref{sec:first} we find the first non-trivial corrections 
\be
\label{slanswer1}
\ba{l}
\dps
R_\alpha(t) = \frac{\delta}{2}\big(\coth \frac{\beta }{2} \cosh t-\sinh t\big) -\frac{\nu \delta}{2}\, \text{csch}\frac{\beta }{2} \cosh \Big((-)^{\alpha+1}\frac{\beta }{2}+t-y\Big)+\cO(\nu^2, \delta^3)\;,
%\\
%\\
%\dps
%R_2(t) = \frac{\delta}{2}\big(\coth \frac{\beta }{2} \cosh t-\sinh t\big) -\frac{\nu\delta}{2} \text{csch}\frac{\beta }{2} \cosh \big(-\frac{\beta }{2}+t-y\big)+\cO(\nu^2, \delta^3)\;,
\ea
\ee
and 
\be
\label{slanswer2}
\ba{l}
\dps
\tilde s_\alpha = 1+\frac{\delta^2}{8}  \text{csch}^2\,\frac{\beta }{2} + \frac{\nu \delta^2}{2} \frac{\cosh (y-\delta_{\alpha,1}\beta )}{1- \cosh\beta}+\cO(\nu^2, \delta^4)\;.
\ea
\ee
Momenta $s_1$ and $s_2$ can be calculated  using the conservation condition  \eqref{s1s2}. Substituting the above expansions into the total action \eqref{tot} and evaluating the integrals \eqref{length_segments} we find that  
\be
\label{banswer}
L_{dual} = \tilde\epsilon \beta  -\frac{\tilde\epsilon \delta^2}{4} \coth \frac{\beta }{2}+ \frac{\tilde\epsilon\nu \delta^2}{2} \left(\sinh y-\coth\frac{\beta }{2} \cosh y\right)+ \cO(\nu^2, \delta^3)\;,
\ee
where the leading contribution is the known length of the 1-point graph, terms $\cO(\nu\delta)$ can be shown to be absent, while the first non-trivial correction is $\cO(\nu\delta^2)$. This decomposition is consistent  with what we would expect from the boundary side, cf.  \eqref{bloook}.  
 
Using the map \eqref{map} which in our case takes the form $\beta = -\log q$ and $y = -\log x$ we find that the total length \eqref{banswer} goes to 
\be
L_{dual}: = - \tilde \epsilon \log q - \frac{\epsilon\delta^2}{4}\frac{1+q}{1- q}+ \frac{\tilde\epsilon\nu \delta^2}{2}\frac{q+x^2}{x-q x}+ \cO(\nu^2, \delta^3)\;.
\ee
Adding the thermal AdS term $S_{thermal} = \frac{1}{4} \log q$ and expanding in the modular parameter $q$ we find that up to unimportant additive constant $S_{thermal}+L_{dual}(\beta,y) = -f^{\tilde\epsilon,\delta,\nu}(q, x)$ , where the perturbative block is given by \eqref{bloook}, or, in a more refined form by  \eqref{g1}. In this way we reproduce the identification formula \eqref{identification}.

\subsection{Double leg expansion } 

We will now move on to consider the second perturbation theory. In this case we assume that conformal dimensions are constrained by \eqref{constraints2}. Most of the steps are similar to that of the previous section so here we only write down the resulting expressions. A more detailed analysis is relegated to Appendix \bref{sec:first}. 

The radial functions of two loop segments and the corresponding momenta are 
\be
\label{jul201}
R_\alpha(t) = \delta\,\text{csch}\frac{\beta }{2} \,\cosh \left(\frac{y }{2} - \delta_{\alpha,1}\frac{\beta}{2}\right) \cosh \left(t-\frac{y }{2}- \delta_{\alpha,0}\frac{\beta}{2}\right) +\cO(\delta^2)\;,
\ee
and 
\be
s_\alpha = \frac{\delta}{2} \left(\coth \frac{\beta }{2} \sinh y-\cosh y\right) +\cO(\nu^2, \delta^2)\;, 
\ee
\be
\label{jul203}
\ba{l}
\dps
\tilde s_\alpha = 1+\frac{\delta^2}{8}  \text{csch}^2\,\frac{\beta }{2} + \frac{\nu \delta^2}{2} \frac{\cosh (y-\delta_{\alpha,1}\beta )}{1- \cosh\beta}+\cO(\nu^2, \delta^4)\;.
\ea
\ee
We note that in the first order the external momenta coincide. Moreover, both $s_{\alpha}\neq 0$, i.e. contrary to the 1-point case the legs are curved, cf. Fig. \bref{annulus}. It follows that the loop segments have different momenta and, therefore, the whole loop is inhomogeneously stretched despite that the  loop dimensions are equal.    

Substituting the above expansions into \eqref{tot0} and \eqref{length_segments} we obtain  that  in the first non-trivial order the total action is given by,
\be
\label{ltotd}
L_{dual}  = \tilde\epsilon\beta- \tilde\epsilon \delta^2 \,\text{csch}\frac{\beta }{2} \cosh \frac{y}{2} \cosh\frac{\beta -y}{2}+ \cO(\delta^3)\;,
\ee
and, finally, using the map \eqref{map} we arrive at 
\be
L_{dual}= -\tilde\epsilon \log q - \frac{\tilde\epsilon \delta^2}{2} \frac{(1+x) (q+x)}{x(1- q)}+\cO(\delta^3)\;.
\ee
Expanding in the modular parameter $q$ we find that modulo an additive constant $S_{thermal}+L_{dual}(\beta,y) = -f^{\delta,\tilde\epsilon}(q, x)$, where the perturbative block is given by \eqref{classblockhigh}, or, in a more refined form by  \eqref{g2}. This proves the identification formulas \eqref{identification} and \eqref{map}.

\section{Concluding remarks}
\label{sec:discussion}

We considered the semiclassical holographic duality in the case of Virasoro CFTs on the torus. We  studied classical torus blocks from various perspectives, in particular, we showed that they are dual to geodesic networks stretched in the thermal AdS bulk space, Fig. \bref{duality}. In the $n$-point case we formulated the system of differential and algebraic equations that govern the dual network. Using various approximation schemes we explicitly solved the system in the 2-point case. 

Our analysis in this paper was mainly focused on the perturbative classical torus blocks in the $s$-channel. We explicitly showed that the perturbative classical blocks within one or another perturbation theory are  equal to the classical global blocks in the limit of large dimensions. Nevertheless, we also analyzed different forms of the $t$-channel torus blocks: quantum, global and classical. In particular, we calculated  the perturbative classical global block in the case of equal external dimensions and heavy loop channel. In this respect our results prove one of the main conjectures of \cite{Kraus:2017ezw} that global blocks are related to the perturbative classical blocks.

In this paper the geodesic networks are described using the first order formulation which results from partially integrating the standard second-order geodesic equations. In this case the integration  constants arising when going to the first order formulation are conserved momenta. In the sphere CFT case, it turns out that momenta are holographically related to the accessory parameters of the monodromy method and this observation is instrumental in proving the holographic  duality between classical blocks and geodesic lengths in the $n$-point case \cite{Hijano:2015rla,Alkalaev:2015lca,Alkalaev:2016rjl}. In this respect, the monodromy method on the torus \cite{Marshakov:2010fx,Menotti:2012wq,KashaniPoor:2012wb,Piatek:2013ifa} which is essentially based on Virasoro symmetry provides  an interesting possibility to go beyond  the $sl(2)$ Casimir equation analysis of \cite{Kraus:2017ezw}. 

The study of the large-$c$ torus CFT and semiclassical duality can be extended in many ways. For example, conformal blocks considered in this paper were  defined in two channels only. The next natural step would be  to generalize to other channels and  identify the corresponding bulk backgrounds that is intimately related to perturbation schemes we use.  Also, it would be interesting to analyze the  torus correlation functions with heavy insertions in the Liouville theory along the lines of Refs. \cite{Harlow:2011ny,Balasubramanian:2017fan}.

%Check the papers by Beccaria and Co, where they cite our first paper on 1-pt torus block.   

\vspace{7mm}
\noindent \textbf{Acknowledgements.} We are grateful to A. Bernamonti, F. Galli, M. Kalenkov and E. Skvortsov for useful discussions. K.A. is grateful to E. Joung for hospitality at Kyung Hee University during the workshop "New ideas on higher spin gravity and holography". K.A. also thanks the Galileo Galilei Institute for Theoretical Physics (GGI) for  hospitality and INFN  within the program "New Developments in AdS3/CFT2 Holography".
The work of K.A. was supported by the Russian Science Foundation grant 14-42-00047. The work of V.B. was supported by the Foundation for the advancement of theoretical physics ``BASIS''.

\appendix

\section{The $s$-channel torus block}
\label{app:matr}

For illustrative purposes, using the general formula \eqref{necklace} we explicitly find the block coefficients in the linear order in the modular parameter,
\be
\label{blockexp}
\cV^{\Delta_{1,2},\tilde\Delta_{1,2}}_{c}(q, z_{1,2}) = \cA^{\Delta_{1,2},\tilde\Delta_{1,2}}_{c}(z_{1,2})+ q \,\cB^{\Delta_{1,2},\tilde\Delta_{1,2}}_{c}(z_{1,2}) + \cO(q^2)\;,
\ee
with coefficients   
\be
\label{ABtotal1}
\cA^{\Delta_{1,2},\tilde\Delta_{1,2}}_{c}(z_{1,2}) = \sum_{m=0}^\infty\;\sum_{|S|=|T|=m} 
\frac{\langle \tilde \Delta_1|\phi_1(z_1)|S,\tilde\Delta_2\rangle}{\langle \tilde \Delta_1|\phi_1(z_1)|\tilde\Delta_2\rangle}\; B^{S|T}_2 \; \frac{
\langle \tilde \Delta_2, T |\phi_2(z_2)|\tilde\Delta_1\rangle}{
\langle \tilde \Delta_2 |\phi_2(z_2)|\tilde\Delta_1\rangle}\;,
\ee
\be
\label{ABtotal2}
\cB^{\Delta_{1,2},\tilde\Delta_{1,2}}_{c}(z_{1,2}) = \sum_{m=0}^\infty \;\sum_{|S|=|T|=m}  
\frac{1}{2 \tilde \Delta_1}\frac{\langle \tilde \Delta_1| L_1 \phi_1(z_1)|S,\tilde\Delta_2\rangle}{\langle \tilde \Delta_1|\phi_1(z_1)|\tilde\Delta_2\rangle}\; B^{S|T}_2\;
\frac{\langle \tilde \Delta_2, T |\phi_2(z_2) L_{-1}|\tilde\Delta_1\rangle}{
\langle \tilde \Delta_2 |\phi_2(z_2)|\tilde\Delta_1\rangle}\;,
\ee
where the Gram matrix on the first $m=0,1,2$ levels in the basis $\{\mathbb{1}, L_{-1}, L_{-2}, L_{-1}^2, ... \}$ is given by  
\be
\label{38}
B{}_{M|N} = 1\;, 
\qquad 
B{}_{M|N} = 2 \tilde \Delta\;, 
\qquad 
B{}_{M|N} =\left(
 \begin{array}{cc}
 \dps\frac{c}{2}+4 \tilde \Delta  & 6 \tilde \Delta  \\
 6 \tilde \Delta  & 4 \tilde \Delta  (2 \tilde \Delta +1) \\
\end{array}
\right)\;.
\ee
In a given order, the  block coefficients can be represented as the  matrix product, 
\be
\label{matrcoef}
\cV = \varkappa\,\tr B_1^{-1}L B_2^{-1} R\;,
\ee 
where $B_{1,2}$ are the Gram matrices, $L$ and $R$ are matrix elements of the primary operators $\phi_m(z_m)$, and  $\varkappa = \varkappa_1 \varkappa_2$ = $z_1^{\Delta_1} z_2^{\Delta_2}\left(z_2/z_1\right)^{\tilde \Delta_1 -\tilde \Delta_2}$, where $\varkappa_1 = 1/\langle \tilde \Delta_1|\phi_1(z_1)|\tilde\Delta_2\rangle$ and $\varkappa_2 = 1/\langle \tilde \Delta_2 |\phi_2(z_2)|\tilde\Delta_1\rangle$.

Let $x = z_2/z_1$. Then, the expansion coefficients up to the second order in $x$ are given by  
\be
\label{A1}
\cA = 1+ \cA_1 x + \cA_2x^2 + ... \;, 
\qquad
\cB = \cB_0x^{-1}+ \cB_1 + \cB_2x + ... \;, 
\ee
where  $\cA_i$ and $\cB_j$ can be read off from the original expressions \eqref{ABtotal1}--\eqref{ABtotal2}. In what follows we calculate them explicitly.  

%\vspace{-5mm}

\paragraph{$\cA$-coefficients.} We easily find the first coefficient  
\be
\label{A1A2}
\cA_1 = \frac{(\Delta_1 +\tilde\Delta_2 -\tilde \Delta_1)(\Delta_2 +\tilde\Delta_2 -\tilde \Delta_1)}{2\tilde\Delta_2}\;.
\ee
Now, we calculate the second coefficient defined as  
$\cA_2 x^2 = \varkappa\, \tr L B^{-1}_2 R$,
where the Gram matrix on the second level is given by \eqref{38}, and matrices $L$ and $R$ are given by 
\be
L = \left(
 \begin{array}{cc}
 \langle \tilde \Delta_1| \phi_1(z_1) L_{-2}|\tilde\Delta_2\rangle   & \dps\langle \tilde \Delta_1| \phi_1(z_1) L_{-1}^2|\tilde\Delta_2\rangle
\end{array}
\right)\;,
\qquad 
R = \left(
 \begin{array}{c}
 \langle \tilde \Delta_2| L_2\,\phi_2(z_2) |\tilde\Delta_1\rangle
  \\ 
 \dps\langle \tilde \Delta_2|L_{1}^2\, \phi_2(z_2) |\tilde\Delta_1\rangle
\end{array}
\right)\;.
\ee
A direct calculation yields 
\be
\label{A9}
\ba{l}
\langle \tilde \Delta_1| \phi_1(z_1) L_{-2}|\tilde\Delta_2\rangle =\varkappa_1^{-1} (\tilde\Delta_2+2\Delta_1-\tilde\Delta_1) z_1^{-2} 
\\
\\

\langle \tilde \Delta_1| \phi_1(z_1) L_{-1}^2|\tilde\Delta_2\rangle =\varkappa_1^{-1} (\tilde\Delta_1 - \Delta_1 -\tilde\Delta_2)(\tilde\Delta_1 - \Delta_1 -\tilde\Delta_2-1) z_1^{-2} 
\\
\\
\langle \tilde \Delta_2| L_2\,\phi_2(z_2) |\tilde\Delta_1\rangle =\varkappa_2^{-1}  (\tilde\Delta_2+2\Delta_2-\tilde\Delta_1)z_2^2
\\
\\
\langle \tilde \Delta_2|L_{1}^2\, \phi_2(z_2) |\tilde\Delta_1\rangle =\varkappa_2^{-1} (\tilde\Delta_2+\Delta_2-\tilde\Delta_1)(\tilde\Delta_2+\Delta_2-\tilde\Delta_1+1)z_2^2 
\ea
\ee
where $\varkappa_{1,2}$ are defined below \eqref{matrcoef}. 
Using \eqref{A9} we can finally assemble the coefficient $\cA_2$  as follows 
\be
\ba{l}
\dps
\cA_2 =\frac{1}{8 c \tilde\Delta_2+4c+8 \tilde\Delta_2 (8 \tilde\Delta_2-5)} \Big[(-\tilde\Delta_1+\tilde\Delta_2+\Delta_2) (-\tilde\Delta_1+\tilde\Delta_2+\Delta_2+1)\times 
\\
\\
\dps
\times
\big(\frac{(c+8 \tilde\Delta_2) (\tilde\Delta_1-\Delta_1-\tilde\Delta_2-1) (\tilde\Delta_1-\Delta_1-\tilde\Delta_2)}{\tilde\Delta_2}+12 (\tilde\Delta_1-2 \Delta_1-\tilde\Delta_2)\big)+
\\
\\
\dps
+4 \big(3 \tilde\Delta_1^2-\tilde\Delta_1 (6 \Delta_1+2 \tilde\Delta_2+1)+(\Delta_1-\tilde\Delta_2) (3 \Delta_1+\tilde\Delta_2-1)\big) (\tilde\Delta_1-\tilde\Delta_2-2 \Delta_2)\Big]\;.
\ea
\ee

\paragraph{$\cB$-coefficients.} Analogously, we find 
\be
\label{B1B2}
\ba{c}
\dps
\cB_0 = \frac{(\tilde \Delta_1 +\Delta_1 -\tilde \Delta_2)(\tilde \Delta_1 +\Delta_2 -\tilde \Delta_2)}{2\tilde\Delta_1} \;,
\\
\\
\dps
\cB_1 = \frac{\left((\tilde\Delta_2+\Delta_1-\tilde\Delta_1)(\tilde\Delta_1+\Delta_1-\tilde\Delta_2-1)+2\tilde\Delta_2\right)\left((\tilde\Delta_1+\Delta_2-\tilde\Delta_2)(\tilde\Delta_2+\Delta_2-\tilde\Delta_1-1)+2\tilde\Delta_1\right)}{4\tilde\Delta_1\tilde\Delta_2} \;.
\ea
\ee
Now, we calculate $\cB_2= \frac{\varkappa}{2\tilde\Delta_1} \tr L B_2^{-1} R$, where new matrices $L$ and $R$ are given by 
\be
L = \left(
 \begin{array}{cc}
 \dps \langle \tilde \Delta_1| L_1\phi_1(z_1) L_{-2}|\tilde\Delta_2\rangle & \langle \tilde \Delta_1| L_1 \phi_1(z_1) L_{-1}^2|\tilde\Delta_2\rangle 
\end{array}
\right)\;,
\qquad 
R = \left(
 \begin{array}{c}
 \dps  \langle \tilde \Delta_2| L_2\,\phi_2(z_2)L_{-1} |\tilde\Delta_1\rangle \\ 
 \langle \tilde \Delta_2|L_{1}^2\, \phi_2(z_2) L_{-1}|\tilde\Delta_1\rangle
\end{array}
\right)
\ee
The matrix elements are given by 
\be
\ba{l}
\langle \tilde \Delta_1| L_1\phi_1(z_1) L_{-2}|\tilde\Delta_2\rangle 
=\varkappa_1^{-1} 
(\tilde\Delta_2-\Delta_1-\tilde\Delta_1)(\tilde\Delta_1-2\Delta_1-\tilde\Delta_2+1) z_1^{-1} 
\\
\\
\langle \tilde \Delta_1| L_1\phi_1(z_1) L_{-1}^2|\tilde\Delta_2\rangle
= \varkappa_1^{-1}
(\tilde\Delta_2 +\Delta_1 -\tilde\Delta_1)\left[(\tilde\Delta_2 + \Delta_1 -\tilde\Delta_1+1)(\tilde\Delta_1+\Delta_1-\tilde\Delta_2-2)+2(2\tilde\Delta_2+1)\right] z_1^{-1} 
\\
\\
\langle \tilde \Delta_2| L_2\,\phi_2(z_2)L_{-1} |\tilde\Delta_1\rangle 
= \varkappa_2^{-1}
(\tilde\Delta_1+\Delta_2-\tilde\Delta_2)(\tilde\Delta_2+2\Delta_2-\tilde\Delta_1-1)z_2
\\
\\
\langle \tilde \Delta_2|L_{1}^2\, \phi_2(z_2)L_{-1} |\tilde\Delta_1\rangle 
= \varkappa_2^{-1}
(\tilde\Delta_2+\Delta_2-\tilde\Delta_1)\left[(\tilde\Delta_2 + \Delta_2 -\tilde\Delta_1+1)(\tilde\Delta_1+\Delta_2-\tilde\Delta_2-2)+2(2\tilde\Delta_2+1)\right]z_2 
\ea
\ee
Finally, we assemble the coefficient $\cB_2$ as follows 
$$
\ba{l}
\dps
\cB_2 =\frac{1}{8 \tilde \Delta_1 (2 c \tilde \Delta_2+c+2 \tilde \Delta_2 (8 \tilde \Delta_2-5))}\bigg[(-\tilde \Delta_1+\tilde \Delta_2+\Delta_2) ((\tilde \Delta_1-\tilde \Delta_2+\Delta_2-2) (-\tilde \Delta_1+\tilde \Delta_2+\Delta_2+1)
\\
\\
\dps
+4 \tilde \Delta_2+2)\big(\frac{(c+8 \tilde \Delta_2) (-\tilde \Delta_1+\Delta_1+\tilde \Delta_2) ((\tilde \Delta_1+\Delta_1-\tilde \Delta_2-2) (-\tilde \Delta_1+\Delta_1+\tilde \Delta_2+1)+4 \tilde \Delta_2+2)}{\tilde \Delta_2}
\ea
$$
\be
\ba{l}
\dps
+12 (\tilde \Delta_1+\Delta_1-\tilde \Delta_2) (\tilde \Delta_1-2 \Delta_1-\tilde \Delta_2+1)\big)+4 \big(3 \tilde \Delta_1^3-\tilde \Delta_1^2 (3 \Delta_1+5 \tilde \Delta_2+7)+
\\
\\
\dps
+\tilde \Delta_1 \left(-3 \Delta_1^2+2 \Delta_1 (\tilde \Delta_2+5)+\tilde \Delta_2^2+6 \tilde \Delta_2+2\right)+3 \Delta_1^3-\Delta_1^2 (5 \tilde \Delta_2+7)+
\\
\\
\dps
+\Delta_1 \left(\tilde \Delta_2^2+6 \tilde \Delta_2+2\right)+\tilde \Delta_2 \left(\tilde \Delta_2^2+\tilde \Delta_2-2\right)\big) (\tilde \Delta_1-\tilde \Delta_2+\Delta_2) (\tilde \Delta_1-\tilde \Delta_2-2 \Delta_2+1)\bigg]\;.
\ea
\ee
We note that  the above calculations can be effectively extended to any order using a recursive technique elaborated in \cite{Kanno:2010kj}.

\section{The $t$-channel torus  block}
\label{app:t}
As an example, in this Appendix we calculate a first few  $t$-channel block coefficients by directly calculating the matrix elements in \eqref{stickman}. The OPE of two primary fields \eqref{OPE} is defined by the descendent operator 
\be
\psi_{\tilde\Delta_2}(z_{1,2}) = \phi_{\tilde\Delta_2}(z_2) 
+ \beta_1 (z_1-z_2)  \phi^{(-1,0)}_{\tilde\Delta_2}(z_{2}) 
+\beta_{21}(z_1-z_2)^2 \phi^{(-1,-1)}_{\tilde\Delta_2}(z_{2})
+\beta_{22}(z_1-z_2)^2 \phi^{(-2,0)}_{\tilde\Delta_2}(z_{2})+ ...  \;,
\ee
where $\phi^{(-1,0)}_{\tilde\Delta_2}(z_{2})  = (L_{-1}\phi_{\tilde\Delta_2})(z_2)$, $\phi^{(-1,-1)}_{\tilde\Delta_2}(z_2)  = (L^2_{-1}\phi_{\tilde\Delta_2})(z_2)$, and $\phi^{(-2,0)}_{\tilde\Delta_2}(z_2)  = (L_{-2}\phi_{\tilde\Delta_2})(z_2)$ are descendants of the primary field $\phi_{\tilde\Delta_2}(z_2)$. The  lowest level $\beta$-coefficients are explicitly known (see, e.g., \cite{ZZbook})
\be
\ba{l}
\dps
\beta_1 = \frac{\tilde\Delta_2+\Delta_1 - \Delta_2}{2\tilde\Delta_2}\;,
\quad
\beta_{21} = \frac{(\tilde\Delta_2+\Delta_1-\Delta_2)(\tilde\Delta_2+\Delta_1-\Delta_2+1)}{4\tilde\Delta_2(2\tilde\Delta_2+1)} - \frac{3}{2(2\tilde\Delta_2+1)} \beta_{22}\;,
\\
\\
\dps
\beta_{22}= \left(\frac{\Delta_1+\Delta_2}{2}+\frac{\tilde\Delta_2(\tilde\Delta_2-1)}{2(2\Delta_2+1)}- \frac{3(\Delta_1-\Delta_2)^2}{2(2\Delta_2+1)}\right)\Big/\left(4\tilde\Delta_2 +\frac{c}{2}- \frac{9\tilde\Delta_2}{2\tilde\Delta_2+1}\right)\;.
\ea
\ee 
Using the Gram matrix expressions \eqref{38} we find in the first order in the modular parameter  
\be
\label{210}
\cV_c^{\Delta_{1,2},\tilde\Delta_{1,2}}(q, z_{1,2}) = \cA_c^{\Delta_{1,2},\tilde\Delta_{1,2}}(q, z_{1,2})+ q \,\cB_c^{\Delta_{1,2},\tilde\Delta_{1,2}}(q, z_{1,2}) + \cO(q^2)\;,
\ee
where  
\be
\label{expAB}
\ba{l}
\dps
\cA_c^{\Delta_{1,2},\tilde\Delta_{1,2}}(q, z_{1,2}) = \varkappa\,  
\langle \tilde \Delta_1 |\psi_{\tilde\Delta_2}(z_{1,2})|\tilde\Delta_1\rangle\;,
\quad 
\cB_c^{\Delta_{1,2},\tilde\Delta_{1,2}}(q, z_{1,2}) = \frac{\varkappa}{2\tilde\Delta_1}\langle \tilde \Delta_1 | L_1 \psi_{\tilde\Delta_2}(z_{1,2}) L_{-1}|\tilde\Delta_1\rangle\;,
\ea
\ee
where $\varkappa^{-1} =\langle \tilde \Delta_1|\phi_{\tilde\Delta_2}(z_2)|\tilde\Delta_1\rangle$. 
Let $w = (z_1-z_2)/z_2$. Then, the expansion coefficients \eqref{expAB} up to the second order in $w$ can be represented as   
\be
\label{A2}
\cA = 1+ \cA_1w + \cA_2w^2 + ... \;, 
\qquad
\cB = \cB_0+\cB_1w+ \cB_1w^2 +  ... \;. 
\ee
The matrix elements are found to be 
$$
\ba{l}
\langle \tilde \Delta_1 |\phi^{(-1,0)}_{\tilde\Delta_2}|\tilde\Delta_1\rangle = -\varkappa^{-1} \tilde\Delta_2 w^{-1} \;,
\quad
\langle \tilde \Delta_1 |\phi^{(-1,-1)}_{\tilde\Delta_2}|\tilde\Delta_1\rangle = \varkappa^{-1}  \tilde\Delta_2(\tilde\Delta_2+1) w^{-2} \;,
\\
\\
\dps
\langle \tilde \Delta_1 |\phi^{(-2,0)}_{\tilde\Delta_2}|\tilde\Delta_1\rangle = \varkappa^{-1} (\tilde\Delta_1+\tilde\Delta_2) w^{-2} \;,
\quad
\langle \tilde \Delta_1 |L_1\phi_{\tilde\Delta_2}L_{-1}|\tilde\Delta_1\rangle =\varkappa^{-1}\left[2\tilde\Delta_1 +\tilde\Delta_2(\tilde\Delta_2-1)\right]\;,
\\
\\
\dps
\langle \tilde \Delta_1 |L_1\phi^{(-1,0)}_{\tilde\Delta_2}L_{-1}|\tilde\Delta_1\rangle =  - \varkappa^{-1}\tilde\Delta_2\left[2\tilde\Delta_1 +\tilde\Delta_2(\tilde\Delta_2-1)\right] w^{-1}\;,
\ea
$$
\be
\ba{l}
\dps
\langle \tilde \Delta_1 |L_1\phi^{(-1,-1)}_{\tilde\Delta_2}L_{-1}|\tilde\Delta_1\rangle =\varkappa^{-1}\tilde\Delta_2(\tilde\Delta_2+1)\left[2\tilde\Delta_1 +\tilde\Delta_2(\tilde\Delta_2-1)\right]w^{-2} \;,
\\
\\
\dps
\langle \tilde \Delta_1 |L_1\phi^{(-2,0)}_{\tilde\Delta_2}L_{-1}|\tilde\Delta_1\rangle = \varkappa^{-1}\left(2\tilde\Delta_1(\tilde\Delta_1+\tilde\Delta_2)+(\tilde\Delta_2+1)\left[(\tilde\Delta_2+2)(\tilde\Delta_1+\tilde\Delta_2)-3\tilde\Delta_2\right]\right)w^{-2}\;.
\ea
\ee
For the sake of simplicity, we represent the coefficients in terms of the $\beta$-coefficients as follows  
\be
\ba{l}
\dps
\cA_1 = -\tilde\Delta_2\,\beta_1\;,
\qquad
\cA_2 =  \tilde\Delta_2(\tilde\Delta_2+1)\beta_{21} +(\tilde\Delta_1+\tilde\Delta_2)\beta_{22}\;,
\ea
\ee
and
$$
\dps
\cB_0 = \frac{2\tilde\Delta_1 +\tilde\Delta_2(\tilde\Delta_2-1)}{2\tilde\Delta_1}\;,
\qquad
\cB_1 = \frac{- \tilde\Delta_2\left(2\tilde\Delta_1 +\tilde\Delta_2(\tilde\Delta_2-1)\right) \beta_1}{2\tilde\Delta_1}\;,
$$
\be
\ba{l}
\dps
\cB_2 = \frac{1}{2\tilde\Delta_1}\Big(\tilde\Delta_2(\tilde\Delta_2+1)\left[2\tilde\Delta_1 +\tilde\Delta_2(\tilde\Delta_2-1)\right] \beta_{21}+
\\
\\
\hspace{4cm}  +\left(2\tilde\Delta_1(\tilde\Delta_1+\tilde\Delta_2)+(\tilde\Delta_2+1)\left[(\tilde\Delta_2+2)(\tilde\Delta_1+\tilde\Delta_2)-3\tilde\Delta_2\right]\right)\beta_{22}\Big)\;.
\ea
\ee

\section{Combinatorial representation of the $s$-channel block }
\label{sec:AGT}

Here we explicitly elaborate the combinatorial AGT \cite{Alday:2009aq} representation  of the  torus conformal multi-point blocks. First, we set convenient Liouville-like parametrization. Instead of central charge $c$ and conformal dimensions 
$\Delta_k$
we will use parameters $b$ and $p_k$ according to 
\be
c=1+6Q^2\;,
\qquad
Q=b+\frac{1}{b}\;, 
\qquad
\Delta_k=\frac{Q^2}{4}-p_k\;.
\ee
Let $F(\bf{q})$, where  ${\bf q}=(q_1,...,q_N)$, be the conformal block  of $N$ primary fields $\phi_{\Delta_k}(z_k)$, $k=1,...,N$, on the torus with the modular parameter $q$, associated with the (left) diagram on Fig. \bref{duality} and with the  intermediate channel parameters between two external lines $p_k$ and $p_{k+1}$ being $\tilde{p}_k$.  
The modular parameter $q$ and the holomorphic coordinates $z_k$ are expressed in terms of $\{q_i\}$  via 
\begin{equation}
\label{zq}
q=\prod_{i=1}^N q_i\;,
\qquad
q_{k}=\frac{z_{k+1}}{z_k}\;,
\qquad 
k=1,..., N-1\;.
\end{equation} 
According to AGT the Virasoro conformal blocks obey the following factorization property 
\be
\label{AGT-HZ}
F({\bf q})=F^H({\bf q})   \mathcal{Z}({\bf q})\;,
\ee
where $F^H(\bf{q})$ is the $N$-point conformal block of  Heisenberg primaries  $\tilde{\phi}_{k}(z_k)$ associated with the same diagram. The Heisenberg algebra is defined by 
\be\label{comm_aa}
[a_n, a_m]=\frac{n}{2}\delta_{n+m,0}\;,
\qquad
n,m \in \mathbb{Z}\;,
\ee
and $\tilde{\phi}_{k}(z_k)$ are related to $\phi_{\Delta_k}(z_k)$ by the requirement that
\be\label{comm_aphi}
[a_{\pm n},\tilde{\phi}_k(z)]=(p_k\pm \frac{Q}{2}) z^{\pm n} \tilde{\phi}_k(z)\;, 
\qquad
n \in \mathbb{N}\;,
\ee
and $p_k$ is the  conformal parameter of the corresponding Virasoro primary field $\phi_{\Delta_k}(z_k)$.

Using commutation relations~\eqref{comm_aa}, \eqref{comm_aphi} and diagonal form of the Heisenberg Gram matrix the Heisenberg conformal block $F^H({\bf q})$ coefficients can be computed order by order.\footnote{For the analogous consideration on the sphere see \cite{ Belavin:2011js}. Alternative considerations are discussed in \cite{Marshakov:2009gs, Mironov:2009dr}.}
For example, on the two lower levels we find 
\begin{align}\label{1-contr}
\sim q_i:& \quad M^{11} \langle \tilde{\phi}_{\tilde{p}_{i}}|\tilde{\phi}_{p_i}  a_{-1}|\tilde{\phi}_{\tilde{p}_{i+1}}\rangle \langle \tilde{\phi}_{\tilde{p}_{i+1}}|a_{1}\tilde{\phi}_{p_{i+1}}  |\tilde{\phi}_{\tilde{p}_{i+2}}\rangle=2 (p_{i}-\frac{Q}{2})(p_{i+1}+\frac{Q}{2})\;,
\end{align}
and
\begin{align}\label{2-contr}
\sim q_i q_{i+1}:& \quad \left(M^{11}\right)^2
\langle \tilde{\phi}_{\tilde{p}_{i}}|a_1\tilde{\phi}_{p_i}  a_{-1}|\tilde{\phi}_{\tilde{p}_{i+1}}\rangle \langle \tilde{\phi}_{\tilde{p}_{i+1}}|a_{1}\tilde{\phi}_{p_{i+1}}  a_{-1}|\tilde{\phi}_{\tilde{p}_{i+2}}\rangle=\nonumber\\
&=4 \bigg[(p_i+\frac{Q}{2}) \langle \tilde{\phi}_{\tilde{p}_i}|\tilde{\phi}_{p_i}(1)  a_{-1}|\tilde{\phi}_{\tilde{p}_{i+1}}\rangle+\frac{1}{2}\langle \tilde{\phi}_{\tilde{p}_i}|\tilde{\phi}_{p_i}(1)|\tilde{\phi}_{\tilde{p}_{i+1}}\rangle\bigg]\times\bigg[i\rightarrow i+1 \bigg]\nonumber\\
&=\bigg[2 (p_i-\frac{Q}{2})(p_i+\frac{Q}{2})+1\bigg]\bigg[2 (p_{i+1}-\frac{Q}{2})(p_{i+1}+\frac{Q}{2})+1\bigg]\;,
\end{align}
where $M^{11}$ stands for the inverse (first-level block) Gram matrix and the normalization $\langle \tilde{\phi}_{\tilde{p}_i}|\tilde{\phi}_{p_j}|\tilde{\phi}_{\tilde{p}_{k}}\rangle=1$ is taken into account. 

In the case of 2-point $s$-channel torus Heisenberg block $F^H_{2pt}({\bf q})$ which contributes  to \eqref{AGT-HZ},  we find up to forth degree in $q_i$ the following expression
\begin{align}
\prod_{n=0}^{\infty} \big(1-q^n q_1\big)^{\alpha_{1,2}} \big(1-q^n q_2\big)^{\alpha_{2,1}}  
\big(1-q^{n+1}\big)^{\alpha_{1,1}+\alpha_{2,2}} \;,
\end{align}
where $\frac{w}{z}=q_1$ and $q= q_1q_2$,  and $\alpha_{i,j}=2(p_i-\frac{Q}{2})(p_j+\frac{Q}{2})$.

The second factor in \eqref{AGT-HZ} has the following combinatorial representation 
\begin{equation}
\label{ZZ}
   \mathcal{Z}({\bf q})=
    \sum_{k_i\in \mathbb{N}_{0}}q_{1}^{k_{1}}q_{2}^{k_{2}}\dots q_{N}^{k_{N}}\,\mathcal{Z}_{k}(\Delta,\tilde{\Delta},c),
\end{equation}
where $\mathbb{N}_{0}$ stands for non-negative integers,  and
\begin{multline}
\label{Zvac}
   \mathcal{Z}_{k}(\Delta,\tilde{\Delta},c)=\!\!\!\!
\sum_{\vec{\lambda}_{1},\dots,\vec{\lambda}_{N}}\!\!\!\!\!\frac{Z(p_{1}|\tilde{p}_0,\lambda_0;\tilde{p}_{1},\vec{\lambda}_{1})Z(p_{2}|\tilde{p}_{1},\vec{\lambda}_{1};\tilde{p}_{2},\vec{\lambda}_{2})
   \cdots 
Z(p_{N}|\tilde{p}_{N-1},\vec{\lambda}_{N-1};\tilde{p}_N,\lambda_{N})}
{Z(\frac{Q}{2}|\tilde{p}_1,\vec{\lambda}_1;\tilde{p}_1,\vec{\lambda}_1) \cdots  
Z(\frac{Q}{2}|\tilde{p}_{N},\vec{\lambda}_{N};\tilde{p}_{N},\vec{\lambda}_{N})}.
\end{multline}
Here, the sum 
goes over $N$ pairs of Young tableaux $\vec\lambda_j =(\lambda_j^{(1)}, \lambda_j^{(2)})$\footnote{The pairs of diagrams label the elements of the orthogonal basis in the representation space of the composite algebra $H\otimes Vir$ \cite{Alba:2010qc}.}  with the total number of cells $|\vec{\lambda}_{j}|\equiv |\lambda_j^{(1)}|+|\lambda_j^{(2)}|=k_{j}$ and we identify $\tilde{p}_0=\tilde{p}_N$, $\lambda_0=\lambda_N$.  The explicit form of  $Z$  reads
\begin{align} 
\label{ZZ}
    Z(P''|P',\vec{\mu};P,\vec{\lambda})=
\prod_{i,j=1}^{2}&\,\bigg[
    \prod_{s\in \lambda_{i}}\left(P''-E_{\lambda_{i},\mu_{j}}\bigl((-1)^j P'-(-1)^i P\bigl|s\bigr)+\frac Q2\right)\nonumber\\
&\,\,\times    \prod_{t\in \mu_{j}}\left(P''+E_{\mu_{j},\lambda_{i}}\bigl((-1)^i P-(-1)^j P'\bigl|t\bigr)-\frac Q2\right)\bigg],  
\end{align}
where
\begin{equation}
\label{E-def}
    E_{\lambda,\mu}\bigl(x\bigl|s\bigr)=x-b\,l_{\mu}(s)+b^{-1}(a_{\lambda}(s)+1).
\end{equation}
For a cell $s=(m,n)$ such that $m$ and $n$ label a respective row  and a column,
the arm-length function $a_{\lambda}(s) = (\lambda)_m-n$ and 
the leg-length function $l_{\lambda}(s) = (\lambda)^T_n - m$, where  
$(\lambda)_m$ is the length of  $m$-th row of the Young tableau $\lambda$,
and  $(\lambda)^T_n$ the height of the $n$-th column, where $(\lambda)^T$ stands for the transposed  Young tableau. 

 Using the final combinatorial expression for the 2-point $s$-channel torus block we find the block coefficients reproducing those in Appendix \bref{app:matr}.

\section{Perturbative  classical $s$-channel torus blocks}
\label{app:block}

Using the combinatorial representation of the 2-point torus block elaborated in Appendix \bref{sec:AGT} we find  the perturbative classical $s$-channel block:

\paragraph{Superlight expansion.}   
\begin{align}
\label{bloook}
&f^{\tilde\epsilon,\delta,\nu}(q, x)=(\tilde\epsilon - 1/4) \log q+ \tilde\epsilon\bigg[\delta^2 \left(\frac{q}{2}+\frac{q^2}{2} + ...\right)+\delta ^4 \left(\frac{q^2}{16}+ ...\right)\bigg]+\nonumber\\
&\qquad+\tilde\epsilon\nu  \bigg[\delta ^2 \left(\frac{q^2}{2 x}+\frac{q x}{2}+\frac{q}{2 x}+\frac{x}{2}+...\right)+
\delta^3 \left(\frac{q^3}{4 x^2}+\frac{q^2}{8 x^2}-\frac{q^2}{2}+\frac{q x^2}{4}-\frac{q}{4}+\frac{x^2}{8}+...\right)+\nonumber\\
&\qquad\quad\delta^4 \left(\frac{q^3}{24 x^3}-\frac{q^2}{8 x}-\frac{q x}{8}+\frac{x^3}{24}+...\right)+\delta ^5 \left(\frac{q^4}{64 x^4}-\frac{q^3}{16 x^2}+\frac{3 q^2}{32}-\frac{q x^2}{16}+\frac{x^4}{64}+...\right)+ ...\bigg]\;,
\end{align}
where the ellipses  denote higher orders in $q,x$ and $\tilde\epsilon,\delta, \nu$.

\paragraph{Double leg expansion.} 
$$
\ba{l}
\dps
f^{\delta,\tilde\epsilon}(q, x) = (\tilde\epsilon - 1/4) \log q+ 
\\
\\
\dps
\hspace{16mm} +\tilde\epsilon\delta^2 \left(q^2 \left(\frac{1}{2 x}+1\right)+q \left(\frac{x}{2}+\frac{1}{2 x}+1\right)+\frac{x}{2}\right)
\\
\\
\dps
\hspace{16mm}+\tilde\epsilon\delta^3 \left(\frac{q^3}{2 x^2}+q^2 \left(\frac{1}{4 x^2}-1\right)+q \left(\frac{x^2}{2}-\frac{1}{2}\right)+\frac{x^2}{4}\right)
\\
\\
\dps
\hspace{16mm}+\tilde\epsilon\delta^4 \left(q^3 \left(\frac{5}{24 x^3}-\frac{1}{4 x^2}\right)+q^2 \left(-\frac{1}{16 x^2}-\frac{3}{8 x}-\frac{1}{8}\right)+q \left(-\frac{x^2}{4}-\frac{3 x}{8}\right)+\frac{5 x^3}{24}-\frac{x^2}{16}\right)
\ea
$$
\be
\label{classblockhigh}
\ba{l}
\dps
\hspace{16mm}+\tilde\epsilon\delta^5 \left(\frac{7 q^4}{32 x^4}+q^3 \left(-\frac{1}{8 x^3}-\frac{3}{8 x^2}\right)+q^2 \left(\frac{1}{8 x}+\frac{5}{16}\right)+q \left(-\frac{3 x^2}{8}+\frac{x}{8}\right)+\frac{7 x^4}{32}-\frac{x^3}{8}\right)

\\
\\
\dps
\hspace{16mm}+\tilde\epsilon\delta^6 \left(-\frac{7 q^4}{32 x^4}+q^3 \left(\frac{1}{48 x^3}+\frac{5}{16 x^2}\right)+\frac{5 q x^2}{16}-\frac{7 x^4}{32}+\frac{x^3}{48}\right)
\\
\\
\dps
\hspace{16mm}+\tilde\epsilon\delta^7 \left(\frac{5 q^4}{64 x^4}-\frac{q^3}{16 x^2}-\frac{q^2}{32}-\frac{q x^2}{16}+\frac{5 x^4}{64}\right)
\\
\\
\dps
\hspace{16mm}+\tilde\epsilon\delta^8 \left(-\frac{5 q^4}{512 x^4}-\frac{5 x^4}{512}\right)+ ...\;,

\ea
\ee
where the ellipses  denote higher orders in $q,x$ and $\tilde\epsilon,\delta$.

\section{Perturbation theory  in the bulk}
\label{sec:first}

\paragraph{Superlight expansion.} Let  $F = \{s_\alpha, \tilde s_\alpha,  \gamma_\alpha, \rho_\alpha, R_\alpha, L_{loop}^\alpha, L_{leg}^\alpha\,|\, \alpha=1,2\}$ be a double series in the deformation parameters  $\nu$ and $\delta$, 
\be
\label{jun31}
F =  \sum_{m,n=0}^\infty  F_{[mn]} \,\nu^m\,\delta^n \;. 
\ee

Using the general formulas \eqref{general} we solve the conservation conditions  as 
\be
\label{s1s2}
s_1 = \frac{\tilde s_2 - \tilde s_1}{\delta}\;, 
\qquad
s_2 = \frac{s_1}{\nu}\;,
\ee
and 
\be
\label{rho1}
\rho_1 = \left[\tilde{s}_1 \tilde{s}_2+\frac{\left(\tilde{s}_1-\tilde{s}_2\right){}^2}{\delta^2}+\frac{\delta^2}{4}-1\right]^{1/2}\left(1-\frac{\delta^2}{4}\right)^{-1/2}\;, 
\qquad
\rho_2 = \rho_1\big|_{\delta\to \delta\nu}\;.
\ee

In what follows we discuss several peculiarities of geodesic equations not seen previously in the 1-point torus case and in the $n$-point sphere case. The main observation is that the angle (time) positions must be known explicitly because they are the integration limits in the loop segment integrals.    

\paragraph{(I)} First of all, we note that defining the radial vertex positions as $\rho_\alpha = R_\alpha(\gamma_\alpha)$ and then decomposing in $\nu$  we find in the lowest orders relations like 
\be
\label{D1}
\rho_{\alpha[0]} = R_{\alpha[0]}(\gamma_{\alpha[0]})\;,
\qquad \rho_{\alpha[1]} = R_{\alpha[1]}(\gamma_{\alpha[0]})+ \gamma_{\alpha[1]} \dot R_{\alpha[0]}(\gamma_{\alpha[0]})\;,\qquad \ldots
\ee 
where the dot denotes a time derivative.  Such decompositions do not arise in the 1-point case because the external leg is radial, and, therefore, the angle position is fixed and coincides with the boundary attachment position.

When defining the superlight expansions \eqref{jun31} we have to assume that $\nu$-corrections start with $\delta^1$. It follows that setting  $\delta=0$ we are guaranteed that all $\nu$-dependent terms vanish, i.e. once the first leg is switched off than the second leg is vanished identically. Hence, 
\be
\label{consist}
R_{\alpha[n0]}(t) = 0\;,
\quad \tilde s_{\alpha[n0]} = 0\;, \quad s_{\alpha[n0]}=0\;, \quad   \rho_{\alpha[n0]}= 0 \;, \quad \gamma_{\alpha[n0]} = 0\;, \qquad n\geq 1\;.
\ee
Using these formulas we find that in the first non-trivial orders 
\be
\label{boundnew}
\rho_{\alpha[00]} = 0\;, 
\qquad
\rho_{\alpha[01]} = R_{\alpha[01]}(\gamma_{\alpha[00]})\;,
\qquad 
\rho_{\alpha[10]} = 0\;, 
\qquad
\rho_{\alpha[11]} = R_{\alpha[11]}(\gamma_{\alpha[00]})\;.
\ee
In other words, the radial vertex positions are defined by radial functions at the seed angle values, $\gamma_{\alpha[00]} \equiv  y_\alpha = \{0,y\}$. This property holds only in the first non-trivial order, while in higher orders we will have to use the general expansion formulas like \eqref{D1}. Note that relations \eqref{boundnew} considerably simplify the analysis of the evolution equation and   the vertex boundary conditions.

To find how the angles flow with the deformation parameters we use the interval equation \eqref{angvert}. Assuming that the seed solution is given by \eqref{srho} we find that  
\be
\label{D2}
\gamma_1 =  \frac{\tilde{s}_{1[12]}-\tilde{s}_{2[12]}}{2}\, \nu \delta + ...\;,
\qquad
\gamma_2 = y+ \frac{\tilde{s}_{1[12]}-\tilde{s}_{2[12]}}{2}\, \delta + \frac{\tilde{s}_{1[22]}-\tilde{s}_{2[22]}}{2}\, \nu \delta+ ...\;,
\ee
where the ellipses denote higher order terms. The above relations say that the angle (time) positions of the boundary attachment points and vertex points are generally different: in the zeroth order they coincide, but switching the interaction on they are starting to fall apart.

\paragraph{(II)} The parameter flow of the angle positions \eqref{D2} makes  integrating along the loop segments \eqref{length_segments}  technically involved  because the integration limits are now double power series in $\delta$ and $\nu$. Indeed, for a given integral  \eqref{length_segments} there is the standard expansion formula  
\be
\label{decF}
\ba{l}
\dps
\int_{\gamma_1(\nu, \delta)}^{\gamma_2(\nu, \delta)} F(x, \nu, \delta) dx  = \int_{\gamma_1(0, \delta)}^{\gamma_2(0, \delta)} F(x, 0, \delta) dx + 
\\
\\
\dps
\hspace{27mm}+\nu \left[\int_{\gamma_1(0, \delta)}^{\gamma_2(0, \delta)} \partial_\nu F(x, 0, \delta) dt+F(\gamma_\alpha(0, \delta),0, \delta)\,\partial_\nu \gamma_\alpha(0, \delta)\Big|_{\alpha=1}^{\alpha=2} \right]+ \cO(\nu^2)\;,
\ea
\ee    
where each term is to be further expanded in powers of  $\delta$. We can show that in the first non-trivial orders the right-hand said of \eqref{decF} reduces to the first and second terms, i.e. terms $\partial \gamma_\alpha$ do not contribute. Analogously, expanding the remainder in $\delta$ we find that corrections in $\gamma_\alpha$ do not contribute as well. In other words, in the first non-trivial order in $\nu$ and $\delta$ we can assume that the integration limits are of the zeroth order, $\gamma_\alpha \approx \gamma_{\alpha[0,0]}$, while the integrands are expanded in the standard fashion. In higher orders this is not generally true.

\paragraph{(III)} Following the general scheme described in Section \bref{sec:dual} we shall find the general solution of the evolution equations. To this end we will solve the system of algebraic equations involving the vertex positions, integration constants, and momenta. Finally, the resulting radial functions $R_{m}(t)$ are used to integrate along the loop.

Let us decompose \eqref{rho1} in $\nu$ and then in $\delta$. To avoid poles in $\nu$ we  set $\tilde{s}_{\alpha[11]}=0$ (additionally to \eqref{consist}).
In the lowest orders we find that radial vertex and momentum corrections are related as 
\be
\label{mo1}
\rho_{1[01]}= \frac{1}{2}\sqrt{8 \tilde{s}_{1[02]}+1}\;,
\qquad 
\rho_{2[01]}=\sqrt{\left(\tilde{s}_{1[12]}-\tilde{s}_{2[12]}\right){}^2+2 \tilde{s}_{1[02]}}\;,
\ee
\be
\label{mo2}
\rho_{1[11]}=\frac{\tilde{s}_{1[12]}+\tilde{s}_{2[12]}}{\sqrt{8 \tilde{s}_{1[02]}+1}}\;,
\qquad 
\rho_{2[11]}=\frac{\tilde{s}_{1[12]}+\tilde{s}_{2[12]}+2 \left(\tilde{s}_{1[12]}-\tilde{s}_{2[12]}\right) \left(\tilde{s}_{1[22]}-\tilde{s}_{2[22]}\right)}{2 \sqrt{\left(\tilde{s}_{1[12]}-\tilde{s}_{2[12]}\right){}^2+2 \tilde{s}_{1[02]}}}\;.
\ee
To solve the above relations ee note that they depend on three combinations of the first order loop momenta $\tilde{s}_{1[12]}\pm\tilde{s}_{2[12]}$ and  $\tilde{s}_{1[22]}-\tilde{s}_{2[22]}$. 

The first relation in \eqref{mo1} is just the first order  expansion coefficient of \eqref{srho} of the 1-point case. It follows that $\rho_{1[01]}$ and $\tilde{s}_{1[02]}$ read off from \eqref{seed1}--\eqref{seed2} satisfy this relation. The second equation allows us to fix $\tilde{s}_{1[22]}-\tilde{s}_{2[22]}$. Indeed, from \eqref{boundnew} we know that $\rho_{2[01]} = R_{2[01]}(y)$. Recalling  \eqref{seed1}--\eqref{seed2} we find that 
\be
\label{b}
\tilde{s}_{1[22]}-\tilde{s}_{2[22]} = \frac{1}{2} \big(\coth \frac{\beta }{2} \sinh (y)-\cosh y\big)\;.
\ee

Now, we expand the evolution equation \eqref{evoleq} in powers of $\nu, \delta$. The first non-trivial correction reads 
\be
\dot R_{\alpha[11]}(t)+\frac{\tilde s_{\alpha[12]}- R_{\alpha[01]}(t) R_{\alpha[11]}(t)}{\dot R_{\alpha[01]}(t)} = 0\;,
\ee 
where the dot denotes a time derivative, while  $R_{\alpha[01]}(t)$ is a known function read off from \eqref{seed1}. The general solution is given by 
\be
\label{Rt}
R_{\alpha[11]}(t) = c_\alpha \sinh \left(t-\beta/2\right)+\tilde s_{\alpha[12]} \big(\sinh t-\sinh (t-\beta )\big)\;,
\ee
where $c_\alpha$ are integration constants. We now use this function to solve the boundary conditions \eqref{boundar} that take the form 
\be
\ba{c}
R_{1[11]}(0) = \rho_{1[11]}\;,
\qquad
R_{1[11]}(y) = \rho_{2[11]}\;,
\\

R_{2[11]}(y) = \rho_{2[11]}\;,
\qquad
R_{2[11]}(\beta) = \rho_{1[11]}\;.
\ea
\ee 
This is four linear equations for four variables $c_\alpha, \tilde s_{\alpha[12]}$. Also, we add two equations \eqref{mo2} which relate radial corrections $\rho_{\alpha[11]}$ and loop momenta $\tilde s_{\alpha[12]}$. The full system of six equations on six variables can be solved unambiguously in terms of parameters $\beta$ and $y$. Then, the roots can be used to find the time dependence $R_{\alpha[11]}(t)$ \eqref{Rt}. All in all, we get   
\be
\label{R11a}
R_{\alpha[11]}(t)=-\frac{1}{2} \,\text{csch}\frac{\beta }{2}\,\cosh \Big((-)^{\alpha+1}\frac{\beta }{2}+t-y\Big)\;,
\ee
and 
\be
\label{s1112}
\tilde{s}_{1[12]}=\frac{\cosh (y-\beta )}{2-2 \cosh \beta}\;,
\qquad
\tilde{s}_{2[12]}=\frac{\cosh y}{2-2 \cosh \beta}\;.
\ee
These functions completely fix the form of the 2-point geodesic graph in the first non-trivial order, cf.  \eqref{slanswer1}--\eqref{slanswer2}. The leg momenta can be found using the relation \eqref{s1s2}. 
  
\paragraph{(IV)} Within the superlight expansion the total action \eqref{tot} can be represented as 
\be
L_{dual} = \tilde\epsilon\left[L_{loop}+ \delta\, L^1_{leg} + \nu\,\delta\, L_{leg}^2 \right] = \tilde\epsilon\left[L_{tot[0]}+\nu \delta^2 L_{tot[12]}+ \cO(\nu^2, \delta^3)\right]\;,
\ee
where the leading contribution is the known length of the 1-point graph, terms $\cO(\nu\delta)$ can be shown to be absent, while the first non-trivial correction is given by $L_{tot[1,2]}$. This decomposition agrees with what we would expect from the boundary side, cf.  \eqref{bloook}.  

The zeroth order correction $L_{tot[0]}$ is given by  \cite{Alkalaev:2016ptm}
\be
\label{lto}
L_{tot[0]} =\beta + \delta^2 \int_0^{\beta }  dt \left(R_{1[01]}(t)R_{1[01]}(t) -\tilde s_{1[02]}\right) - \delta^2 R_{1[01]}(0)\;,
\ee 
where the first and second terms are loop contributions, while the third term is the leg contribution. Substituting the explicit solution \eqref{seed1}--\eqref{seed2} into \eqref{lto} we get 
\be
\label{ltot0}
L_{tot[0]} =\beta -\frac{1}{4} \coth \frac{\beta }{2}\;.
\ee
To find the first order correction $L_{tot[12]}$ we use \eqref{tot} and \eqref{length_segments} and  show that
\be
L_{tot[12]} = L^{1}_{loop[12]}+L^{2}_{loop[12]}+L^1_{leg[11]}+L^2_{leg[11]}\;,
\ee 
where   
\be
\ba{c}
\dps
L^{m}_{loop[12]} = \int_{0}^y dt \left(2 R_{m[11]}(t) R_{m[0]}(t)-\tilde{s}_{m[12]}\right)\;,
\quad
m=1,2
\\
\\
L^1_{leg[11]} =  - R_{1[11]}(0) \;, 
\qquad
L^2_{leg[11]} = -R_{2[01]}(y) \;.
\ea
\ee
Here, the integrands are explicitly given by \eqref{seed1}--\eqref{seed2} and \eqref{R11a}--\eqref{s1112}. We find 
\be
\label{ltot1}
L_{tot[12]} = \frac{1}{2} \left(\sinh y-\coth\frac{\beta }{2} \cosh y\right)\;.
\ee
Combining \eqref{ltot0} and \eqref{ltot1} we arrive at the final formula \eqref{banswer}.

\paragraph{Double leg expansion.} Let $F = \{\tilde s_\alpha,  \gamma_\alpha, s, \rho, R_\alpha, L_{loop}^\alpha, L_{leg}^\alpha\,|\, \alpha=1,2\}$ be a power  series in the lightness parameter $\delta$, 
\be
\label{jul18}
F =  \sum_{n=0}^\infty  F_{[n]}\,\delta^n \;.
\ee
As the seed solution $\cO(\delta^0)$ we choose the circle going along the constant zeroth radius and radial legs, cf. \eqref{dotr},
\be
\tilde s_{\alpha[0]} = 1\;, \qquad s_{[0]} = 0\;, \qquad R_{\alpha[0]}(t) = 0\;, \qquad \gamma_{\alpha[0]} = y_\alpha\;,
\ee
where $y_1 = 0$ and $y_2 = y$. Moreover, similar to the analysis of \cite{Alkalaev:2016ptm} we assume that  
$\tilde s_{\alpha[2n+1]} = 0$, $
s_{\alpha[2n]} = 0$, $R_{\alpha[2n]}(t) = 0$ 
for $n \in \mathbb{N}_0$. 

Using constraints \eqref{constraints2} along with the identity $\tilde\epsilon \delta^{n} = \epsilon \delta^{n-1}$ we find from \eqref{tot} that the total length is expanded as
\be
\label{totexp}
L_{dual} = \tilde\epsilon\beta+ \tilde\epsilon \sum_{n=2}^\infty\left[L^1_{loop[n]} + L^2_{loop[n]}+ L^1_{leg[n-1]}+ L^2_{leg[n-1]} \right]\,\delta^n\;,
\ee
where we used that in the zeroth approximation the loop is the constant radius circle and the legs are vanishing, i.e. $L^1_{loop[0]} + L^2_{loop[0]} = \tilde\epsilon\beta$ and $L^{\alpha}_{leg[0]}=0$. In particular, the sum in \eqref{totexp} starts with $n=2$ term. Also, $L^\alpha_{loop[2n+1]} = 0$ for $n\in \mathbb{N}_0$. 

Using explicit expressions \eqref{length_segments} and taking into comments {\bf(I)}--{\bf(III)} above  we find that in the first non-trivial order  
\be
\label{L1d}
\ba{l}
\dps
L^1_{loop} = y + \delta^2 \int_0^y dt \left(R_{1[1]}(t)R_{1[1]}(t)-\tilde s_{1[2]}\right) + \cO(\delta^4)\;,
\\
\\
\dps
L^2_{loop} =  (\beta-y) + \delta^2\int_y^{\beta-y} dt \left(R_{2[1]}(t)R_{2[1]}(t)-\tilde s_{2[2]}\right) +\cO(\delta^4)\;,
\ea
\ee
and
\be
\label{L1d3}
L^1_{leg} = L^2_{leg} = -\rho_{[1]} \delta +  \cO(\delta^2)\;.
\ee

Now, the conservation conditions are solved by 
the general formulas \eqref{general} as follows  
\be
\label{rho111}
\rho = \rho_1 = \rho_2 = \left[\tilde{s}_1 \tilde{s}_2+\frac{\left(\tilde{s}_1-\tilde{s}_2\right){}^2}{\delta^2}+\frac{\delta^2}{4}-1\right]^{1/2}\left(1-\frac{\delta^2}{4}\right)^{-1/2},
\quad s\equiv s_1 = s_2 = \frac{\tilde s_2 -\tilde s_1}{\delta}\;,
\ee
that mean the two legs are symmetrically attached to the loop. To find the radial functions $R_{\alpha[1]}$ we expand the evolution equations in $\delta$. In the first non-trivial order we find the following equations
\be
\frac{R^{\alpha[1]}(t)}{d t}-\sqrt{R^{\alpha[1]}(t)^2-2 \tilde s_{\alpha[2]}} =0\quad \text{solved as}\quad R^{\alpha[1]}(t) = \frac{1}{2} \left(2 \tilde{s}_{\alpha[2]} e^{-c_{\alpha}-t}+e^{c_\alpha+t}\right)\;.
\ee
Integration constants $c_\alpha$ can be fixed by the boundary conditions \eqref{boundar} given in the first non-trivial order. Finally, we find that the radial functions read 
\be
\label{R11}
R_{1[1]}(t) = \rho_{[1]} \left(\cosh t-\tanh \frac{\gamma }{2} \sinh t\right)\;,
\qquad
R_{2[1]}(t) = \rho_{[1]} \frac{ e^{\beta +y -t}+ e^t}{e^{\beta }+e^{\gamma }}\;, 
\ee 
where $\rho_{[1]}$ is the vertex correction, and 
\be
\tilde s_{1[2]} = \frac{\rho^2_{[1]}}{\cosh y+1}\;, 
\qquad 
\tilde s_{2[2]} = \frac{2\, e^{\beta +y }\,\rho^2_{[1]}}{\left(e^{\beta }+e^{y }\right)^2}\;.
\ee 
On the other hand, expanding  \eqref{rho111} we find that $\rho_{[1]}$ and $\tilde s_{\alpha[2]}$ are related as 
\be
\rho_{[1]} =  \sqrt{ (\tilde{s}_{1[2]}-\tilde{s}_{2[2]})^2 + (\tilde{s}_{1[2]}+ \tilde{s}_{2[2]})+\frac{1}{4}}\quad\to\quad \rho_{[1]} = \text{csch}\frac{\beta }{2}\cosh \frac{y}{2} \cosh \frac{\beta -y}{2}\;. 
\ee
All in all, we find that the radial functions and momenta are given by \eqref{jul201}
--\eqref{jul203}. Substituting these expressions into \eqref{L1d}--\eqref{L1d3} we obtain \eqref{ltotd}.

\providecommand{\href}[2]{#2}\begingroup\raggedright
\addtolength{\baselineskip}{-3pt} \addtolength{\parskip}{-1pt}

\providecommand{\href}[2]{#2}\begingroup\raggedright\endgroup

%\bibliographystyle{JHEP}
%\bibliography{refs}

%\bibliography{VirasoroRulesBib}

\end{document}